\documentclass[twocolumn,times,trackchanges,tight]{aastex62}
\usepackage{graphicx}
\usepackage{color}
\usepackage{mathrsfs}
\usepackage{multirow}
\usepackage{amsmath}

\def\bhm{M_{\bullet}}
\newcommand{\civ}{\ifmmode {\text{C~{\sc iv}}} \else C~{\sc iv} \fi}
\newcommand{\lya}{\ifmmode {\rm Ly\alpha} \else Ly$\alpha$ \fi}
\def\oiii{\rm [O~{\sc iii}]}
\def\oiiid{\rm [O~{\sc iii}]$\lambda\lambda4959,5007$}
\def\oiiis{\rm [O~{\sc iii}]$\lambda5007$}
\def\feii{\rm Fe~{\sc ii}}
\def\hbeta{H$\beta$}
\def\hgamma{H$\gamma$}

\def\Nv{N {\sc v}}
\def\sunm{M_{\odot}}

\def\ergs {\rm erg\, s^{-1}}
\def\hst{{\it HST}}

\def\kms{{\rm km\,s^{-1}}}
\def\ser{S\'{e}rsic}
\def\asec{$^{\prime\prime}$}
\def\h{\hskip -1.5 mm}

\def\nl{\\}

\newcommand{\sersic}{S\'{e}rsic}

\received{November 7, 2018}
\revised{February 23, 2019}
\accepted{March 3, 2019}

\begin{document}
\title{\bf\large
Kinematics of the Broad-line Region of 3C~273 from a Ten-year Reverberation Mapping Campaign}

\correspondingauthor{Pu Du, Jian-Min Wang}
\email{dupu@ihep.ac.cn, wangjm@ihep.ac.cn}

\author{Zhi-Xiang Zhang}
\affiliation{Key Laboratory for Particle Astrophysics, Institute of High Energy Physics, Chinese Academy of Sciences, 19B Yuquan Road, Beijing 100049, China}
\affiliation{School of Astronomy and Space Science, University of Chinese Academy of Sciences, 19A Yuquan Road, Beijing 100049, China}

\author{Pu Du}
\affiliation{Key Laboratory for Particle Astrophysics, Institute of High Energy Physics, Chinese Academy of Sciences, 19B Yuquan Road, Beijing 100049, China}

\author{Paul S. Smith}
\affiliation{Steward Observatory, University of Arizona, Tucson, AZ 85721, USA}

\author{Yulin Zhao}
\affiliation{Kavli Institute for Astronomy and Astrophysics, Peking University, Beijing 100871, China}
\affiliation{Department of Astronomy, School of Physics, Peking University, Beijing 100871, China}

\author{Chen Hu}
\affiliation{Key Laboratory for Particle Astrophysics, Institute of High Energy Physics, Chinese Academy of Sciences, 19B Yuquan Road, Beijing 100049, China}

\author{Ming Xiao}
\affiliation{Yunnan Observatories, Chinese Academy of Sciences, Kunming 650011, China}
\affiliation{Key Laboratory for Particle Astrophysics, Institute of High Energy Physics, Chinese Academy of Sciences, 19B Yuquan Road, Beijing 100049, China}

\author{Yan-Rong Li}
\affiliation{Key Laboratory for Particle Astrophysics, Institute of High Energy Physics, Chinese Academy of Sciences, 19B Yuquan Road, Beijing 100049, China}

\author{Ying-Ke Huang}
\affiliation{Key Laboratory for Particle Astrophysics, Institute of High Energy Physics, Chinese Academy of Sciences, 19B Yuquan Road, Beijing 100049, China}
\affiliation{School of Astronomy and Space Science, University of Chinese Academy of Sciences, 19A Yuquan Road, Beijing 100049, China}

\author{Kai Wang}
\affiliation{Key Laboratory for Particle Astrophysics, Institute of High Energy Physics, Chinese Academy of Sciences, 19B Yuquan Road, Beijing 100049, China}
\affiliation{School of Astronomy and Space Science, University of Chinese Academy of Sciences, 19A Yuquan Road, Beijing 100049, China}

\author{Jin-Ming Bai}
\affiliation{Yunnan Observatories, Chinese Academy of Sciences, Kunming 650011, China}

\author{Luis C. Ho}
\affiliation{Kavli Institute for Astronomy and Astrophysics, Peking University, Beijing 100871, China}
\affiliation{Department of Astronomy, School of Physics, Peking University, Beijing 100871, China}

\author{Jian-Min Wang}
\affiliation{Key Laboratory for Particle Astrophysics, Institute of High Energy Physics, Chinese Academy of Sciences, 19B Yuquan Road, Beijing 100049, China}
\affiliation{School of Astronomy and Space Science, University of Chinese Academy of Sciences, 19A Yuquan Road, Beijing 100049, China}
\affiliation{National Astronomical Observatories of China, Chinese Academy of Sciences, 20A Datun Road, Beijing 100020, China}

\begin{abstract}
	Despite many decades of study, the kinematics of the broad-line region of
	3C~273 are still poorly understood. We report a new, high signal-to-noise,
	reverberation mapping campaign carried out from November 2008 to March 2018 that
	allows the determination of time lags between emission lines and the
	variable continuum with high precision. The time lag of variations in
	\hbeta\ relative to those of the 5100\AA\ continuum is
	$146.8_{-12.1}^{+8.3}$ days in the rest frame, which agrees very well with the
	Paschen-$\alpha$ region measured by the GRAVITY at The Very Large Telescope 
	Interferometer. The time lag of the
	\hgamma\  emission line is found to be nearly the same as for \hbeta. The
	lag of the \feii\ emission is $322.0_{-57.9}^{+55.5}$ days, longer by a
	factor of $\sim$2 than that of the Balmer lines. The velocity-resolved lag
	measurements of the \hbeta\ line show a complex structure which can be
	possibly explained by a rotation-dominated disk with some
	inflowing radial velocity in the \hbeta-emitting region. 
	Taking the virial factor of $f_{\rm BLR} = 1.3$,  we derive a BH
	mass of $\bhm = 4.1_{-0.4}^{+0.3} \times 10^8 M_{\odot}$ and an accretion
	rate of $9.3\,L_{\rm Edd}\,c^{-2}$ from the \hbeta\ line. The decomposition
	of its \hst\ images yields a host stellar mass of $M_* = 10^{11.3 \pm 0.7}
	M_\odot$,  and  a ratio of $\bhm/M_*\approx 2.0\times 10^{-3}$ 
	in agreement with the Magorrian relation.  In the near future, it is expected 
	to compare the geometrically-thick BLR 
	discovered by the GRAVITY in 3C 273 with its spatially-resolved torus  
	in order to understand the potential connection between the BLR and the torus.
\end{abstract}


\section{Introduction} \label{sec:intro}

After its discovery \citep{Schmidt1963}, 3C~273 has been intensively studied
over the whole range of the electromagnetic spectrum (from radio to
$\gamma$-ray bands) and has become one of the most representative quasars
\citep{Cour1998}, as it shows most of the intriguing features of active
galactic nuclei (AGNs), e.g., the jet and the large flux variations at all
wavelengths. Although 3C~273 is classified as a blazar for its high-energy
radiation above 100 MeV, it shows a prominent big blue bump and strong
emission lines in the UV and optical bands. These coexistent characteristics
make 3C~273 an interesting object \citep{Turler1999}. However, the geometry
and kinematics of its broad-line region (BLR) are far from fully understood.

It has been shown that reverberation mapping (RM; see, e.g.,
\citealt{Bahcall1972,Bland1982,Peterson1993, Peterson1998b,
Peterson2002, Peterson2004, Kaspi2000, Kaspi2007, Bentz2008, Bentz2009,
Denney2009b, Barth2011, Barth2013, Barth2015, Rafter2013, Du2014,
Du2015, Du2016V, Du2018, Du2018W, Wang2014, Shen2016, Jiang2016, Fausnaugh2017,
Grier2017, De2018}) is a reliable way of measuring the masses of black holes (BHs) in
AGNs (but time-consuming for very massive black holes) and investigating the geometry 
and kinematics of their BLRs. In \cite{Kaspi2000}, the light curves of the
H$\alpha$, H$\beta$, and H$\gamma$ emission lines of 3C~273 versus
the variation of its continuum at 5100\AA\ yielded slightly different
time lags, and the cross-correlation functions (CCFs) peak at lags
of $\tau_{\rm H\alpha}\sim 500$ days, $\tau_{\rm H\beta}\sim 380$ days,
and $\tau_{\rm H\gamma}\sim 300$ days, respectively. The other emission
lines, such as \feii, have not been reported yet. The large season gaps
and the low sampling cadence (about 30 nights) of the light curves of
3C~273 in \citet{Kaspi2000} plausibly influence the accuracy of its time
lag measurement, and result in fairly large uncertainties.
Moreover, the velocity-resolved RM, which measures the time lags of
the emission line at different velocities, is gradually adopted to
understand the BLR kinematics and structure in recent years,
and has been applied to more than a dozen AGNs
\citep[e.g.,][]{Bentz2008, Bentz2009, Bentz2010, Denney2009b,
Denney2010, Grier2013, Kollatschny2013, De2015, Du2016VI, Lu2016, Pei2017, Xiao2018}.
But higher quality spectra are needed for velocity-resolved
analysis than for just getting a mean time lag of an emission line.
With the very massive black hole ($10^8\sim10^9M_{\odot}$) in the
center of 3C~273, it varies on a time scale of years (see Figure 3
in \citealt{Kaspi2000}). Thus, a long-term RM campaign with high
calibration precision, which can continue for several years,
is required.

In this paper, we report a new RM campaign of 3C~273 with high quality
which spans about ten years (from Nov 2008 to Mar 2018). In Section
\ref{sec:data}, we describe the observation and data reduction.
The light curve measurement and the inter-calibration of the data
from the different telescopes are provided in Section
\ref{sec:measure}. The time lags of the emission lines and the
velocity-resolved RM result are given in Section \ref{sec:res}.
The BH mass is estimated in Section \ref{sec:dis}, and the host
decomposition of 3C~273 is also provided in this section. We briefly
summarize this paper in Section \ref{sec:sum}. We adopt the
$\Lambda$CDM cosmology with $\Omega_M=0.32$,
$\Omega_{\Lambda}=0.68$, and $H_0=67\ {\rm km\ s^{-1}\ Mpc^{-1}}$
\citep{Planck2018} in this paper, and the luminosity distance
is 782.7 Mpc.

\section{Observations and Data}\label{sec:data}
\subsection{Steward Observatory spectropolarimetric monitoring project}\label{sec:SOPS}

The spectroscopic and photometric data of 3C~273 used in this paper are
mainly derived from the Steward Observatory (SO) spectropolarimetric monitoring
project\footnote{\url{http://james.as.arizona.edu/~psmith/Fermi/}},
which is a long-term optical program  to support the Fermi $\gamma$-ray Space
Telescope.  The project began just after the launch of Fermi in 2008 and
provides almost a decade of spectropolarimetric, photometric and spectroscopic
data for more than 70 blazars \citep{Smith2009}.

The SO campaign utilizes both the 2.3 m Bok Telescope on Kitt Peak
and the 1.54 m Kuiper Telescope on Mt. Bigelow in Arizona.
All of the observations are carried out using the SPOL
spectropolarimeter \citep{Schmidt1992}, which is a versatile,
high-throughput, and low-resolution spectropolarimeter.
A $600\ {\rm mm^{-1}}$\ grating is used, providing a wavelength
coverage of 4000-7550\AA\ and a spectral resolution of
15$\sim$25\AA\ (FWHM of the line spread function) 
depending on the slit width chosen \citep{Smith2009}.
Four slits were used for the the observations of 3C~273: 4.1$''$, 5.1$''$, 
7.6$''$, and 12.7$''$. Primarily, the 7.6$''$-wide slit was used for the 
spectropolarimetric observations, which yields high-S/N spectra of the 
object. The 12.7$''$-wide slit was used for shorter observations of the quasar 
and a calibrated field comparison star to determine the apparent magnitude of
3C~273 in a synthetic Johnson $V$ filter bandpass and thereby calibrate the
spectrophotometry.

There are a total of 374 spectroscopic epochs for 3C~273 up until 2018
March from the SO program. Flux calibration of the spectra is based on
the average sensitivity function derived from multiple observations of
spectrophotometric standard stars (BD+28~4211 and/or G191~B2B) obtained
during each observing campaign (typically about a week in length).
Further night-to-night relative calibration was accomplished by 
convolving the spectra with a standard Johnson $V$ filter bandpass and 
then scaling the spectrum to agree with the $V$-band photometric 
light curve.

\begin{figure}
	\centering
	\includegraphics[width=0.45\textwidth]{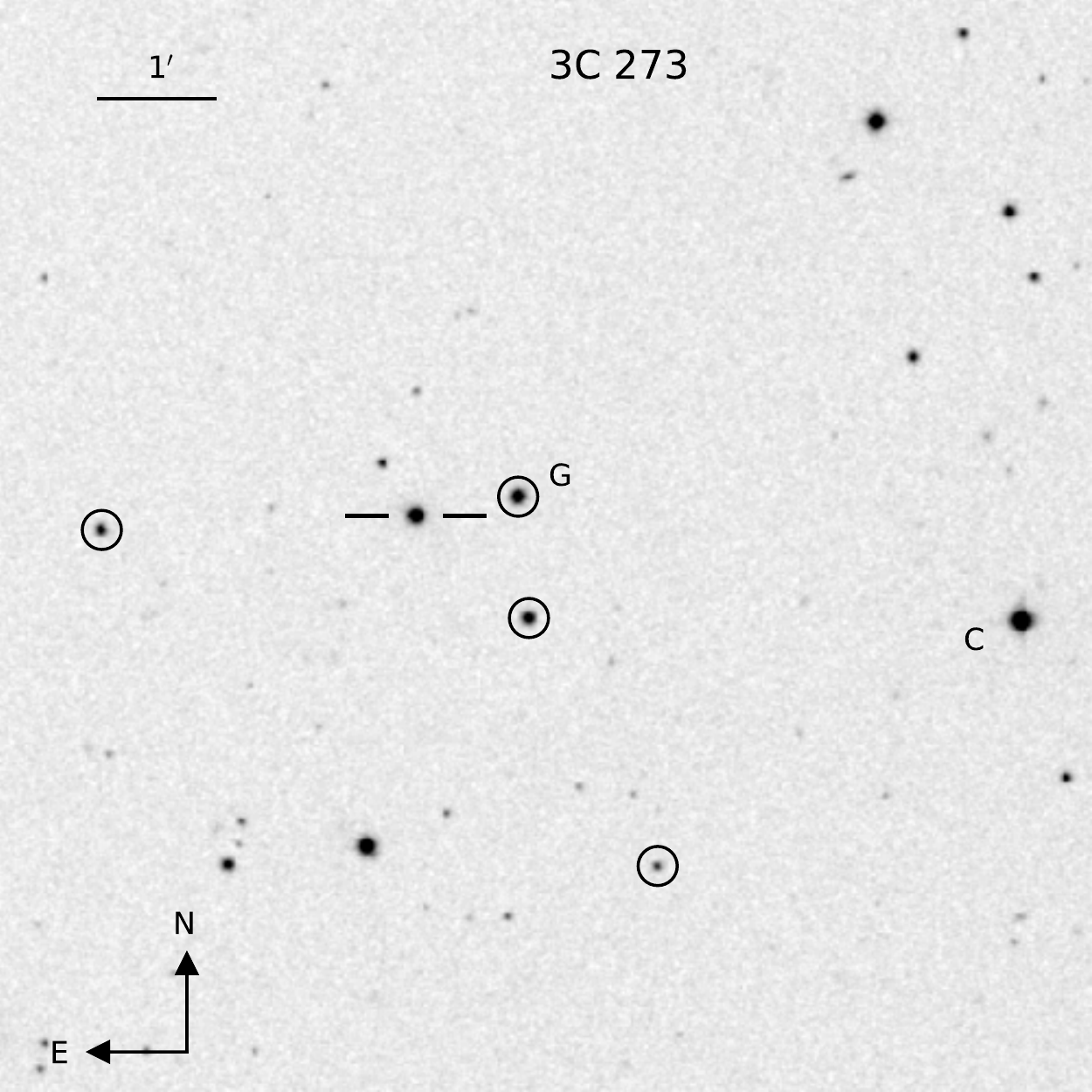}
	\caption{\footnotesize
		The field around 3C~273.
		Star C \citep[$V = 11.87$;][]{Smith1985} is used as the
		spectrophotometric standard star for the SO observation.
		Star G was used as the spectroscopic comparison star for the
		observations from Lijiang Observatory, while stars in circle were
		used to calibrate the Lijiang photometry. Here, we just adopt the same historical 
		designation of ``C'' and ``G'' in \cite{Smith1985}.\label{fig:img}}
\end{figure}

3C~273 has 297 $V$-band photometric epochs, and their magnitudes were
calibrated by differential photometry using 
the star C in Figure \ref{fig:img} (see more details in
\citealt{Smith1985}). 
The $V$-band light curve from SO is shown
in the following Section \ref{sec:intercal}.

For the 374 spectroscopic observations, there are only 291 spectra
that have the photometric observations in the same nights.
They were calibrated using the corresponding photometric magnitudes,
and are used in the following analysis. After combining the exposures
in the same night, there are totally 283 spectroscopic epochs.
The other 83 spectra with no simultaneous photometric observations
are not used in this paper.

It should be noted that  the aperture used in the spectroscopic observation
includes almost all of the starlight from the host galaxy and the host
contribution in the aperture is less than 6\% (estimated from the image
fitting in Section \ref{sec:stellarmass}).  Thus, the night-to-night
calibration should not be significantly influenced by the host contribution in
the different apertures used in the spectroscopy and photometry. The
consistency of the SO and Lijiang light curves in the following Section
\ref{sec:intercal} also supports that.

\begin{deluxetable*}{lcccccccccc}
\tabletypesize{\scriptsize}
	\tablecaption{Continuum and emission-line fluxes of 3C~273 \label{table:lightcurve}}
		\tablehead{
			\multicolumn{6}{c}{Spectra} &
			\colhead{} &
			\multicolumn{3}{c}{Photometry} \\
			\cline{1-6} \cline{8-10}
			\colhead{JD} &
			\colhead{$F_{5100}$} &
			\colhead{$F_{\rm H\beta}$} &
			\colhead{$F_{\rm H\gamma}$} &
			\colhead{$F_{\rm Fe}$} &
			\colhead{Obs.} &
			\colhead{} &
			\colhead{JD ($V$)} &
			\colhead{Mag ($V$)} &
			\colhead{Obs.}
		}
		\startdata
		2454795.02 & $21.75 \pm 0.27$ & $18.17 \pm 0.24$ & $6.93 \pm 0.13$ & $14.95 \pm 0.28$ & S & & 2454795.01 & $12.71 \pm 0.02$ & S \\
		2454800.99 & $22.95 \pm 0.28$ & $18.27 \pm 0.25$ & $6.70 \pm 0.16$ & $15.41 \pm 0.34$ & S & & 2454801.00 & $12.68 \pm 0.02$ & S \\
		2454802.99 & $24.09 \pm 0.27$ & $18.00 \pm 0.23$ & $6.27 \pm 0.12$ & $15.11 \pm 0.27$ & S & & 2454802.99 & $12.66 \pm 0.02$ & S \\
		2454803.97 & $22.28 \pm 0.27$ & $16.92 \pm 0.23$ & $5.90 \pm 0.11$ & $14.41 \pm 0.25$ & S & & 2454804.04 & $12.74 \pm 0.02$ & S \\
		2454828.98 & $20.40 \pm 0.27$ & $17.64 \pm 0.23$ & $6.02 \pm 0.11$ & $14.93 \pm 0.24$ & S & & 2454804.05 & $12.70 \pm 0.02$ & S \\
		\enddata
		\tablecomments{\footnotesize
		$F_{5100}$ is the continuum flux at 5100\AA\ in units
			of ${\rm 10^{-15} erg\,s^{-1} cm^{-2} \AA^{-1}}$. The emission-line flux is given in units of 
            ${\rm 10^{-13} erg\,s^{-1} cm^{-2}}$. Obs. is the observatory, ``S'' $=$ SO, ``L'' $=$ Lijiang, 
			and ``A'' $=$ ASAS-SN. This table is available in its entirety
in a machine-readable form in the on-line journal. A portion is shown here
for guidance regarding its form and content. }
\end{deluxetable*}

\subsection{SEAMBH Campaigns of the Lijiang 2.4m telescope}

As a candidate for super-Eddington accreting massive black 
holes (SEAMBHs), 3C~273 was selected as a target in our SEAMBH campaign
\citep{Du2014, Du2015, Du2016V, Du2018, Wang2014} for cosmology
\citep{Wang2013}. We monitored 3C~273 from December 28, 2016 to
June 13, 2017 using the 2.4 m telescope at the Lijiang Station of the
Yunnan Observatories, Chinese Academy of Sciences. It is mounted with
the Yunnan Faint Object Spectrograph and Camera (YFOSC),
which is a multi-functional instrument both for photometry and spectroscopy.
The YFOSC detector is a ${\rm 2k\times4k}$ back-illuminated CCD with pixel
size of 13.5$\,{\rm \mu m}$, providing a ${\rm 10'\times10'}$ field of view
(${\rm 0''.283}$ per pixel). Grism 14 and a $2''.5$-width slit 
were used, which provide a spectral dispersion of ${\rm 1.74\,\AA\,pixel^{-1}}$ and a resolution 
of $\sim$ 500 km/s \citep{Du2016VI}. The wavelength coverage is
3800--7200\AA, and the wavelength was calibrated by neon and helium lamps.
The spectra were reduced using the standard {\tt IRAF v2.16} procedures.

We adopted the calibration approach based on in-slit comparison star
\citep[e.g.,][]{Maoz1990,Kaspi2000, Du2014, Du2018} rather than the
\oiii-based approach \citep[e.g.,][]{van1992, Peterson2013, Barth2015, Fausnaugh2017}
for the spectroscopic observation. We observed 3C~273 and a nearby
comparison star (star G in Figure \ref{fig:img}) simultaneously by
rotating the long slit, and used the comparison star as a standard
to do the calibration. The fiducial spectrum of the comparison star
is generated by averaging the spectra observed in photometric conditions.
This approach ensures highly accurate relative flux calibration with
a precision of $\sim2\%$ \citep{Du2014, Du2018, Lu2016}.

We also performed photometric observations using the Johnson $V$ filter. 
The light curve of 3C~273 in $V$-band is measured by differential photometry 
using 4 stars (the circled stars in Figure \ref{fig:img}) in the same field.
There are 28 photometric and 27 spectroscopic epochs in Lijiang 
observations, respectively. The average sampling 
interval is $\sim6$ days. The photometric light curve from Lijiang, which has 
been scaled to match the SO $V$-band light
curve, is shown in Section \ref{sec:intercal}.

\subsection{All-Sky Automated Survey for Supernovae project}

For comparison, we also show the $V$-band light curve of 3C~273 from
the All-Sky Automated Survey for Supernovae (ASAS-SN) project
\footnote{\url{http://www.astronomy.ohio-state.edu/asassn/index.shtml}}.
The ASAS-SN project \citep{Shappee2014,Kocha2017} is working toward
imaging the entire sky every night down to $V\sim17$ magnitude,
and spans $\sim2-5$ years with $\sim100-400$ observational epochs for
different objects. 
The ASAS-SN light curve of 3C 273 used here consists of 348 epochs in a 6-year 
period after averaging multiple observations obtained within a night.
The ASAS-SN $V$-band light curve, after scaling to the SO light curve, 
is provided in the following  
Section \ref{sec:intercal}.

\section{measurements}\label{sec:measure}
\subsection{Mean and RMS spectra}\label{sec:meanrms}

	Small wavelength shifts in the spectra, which average to about 1.4\AA\ for
	the SO data and 1.2\AA\ for the Lijiang data, caused by instrument flexure,
	uncertainties in the wavelength calibration, and any mis-centering of the
	object within the slit have been corrected by aligning all spectra to 
	\oiiis. In addition, correction for the Galactic extinction of $A_V=0.057$ 
	has been applied to
	the spectra \citep{Cardelli1989, Schlafly2011}. 

\begin{figure}
	\centering
	\includegraphics[width=0.48\textwidth]{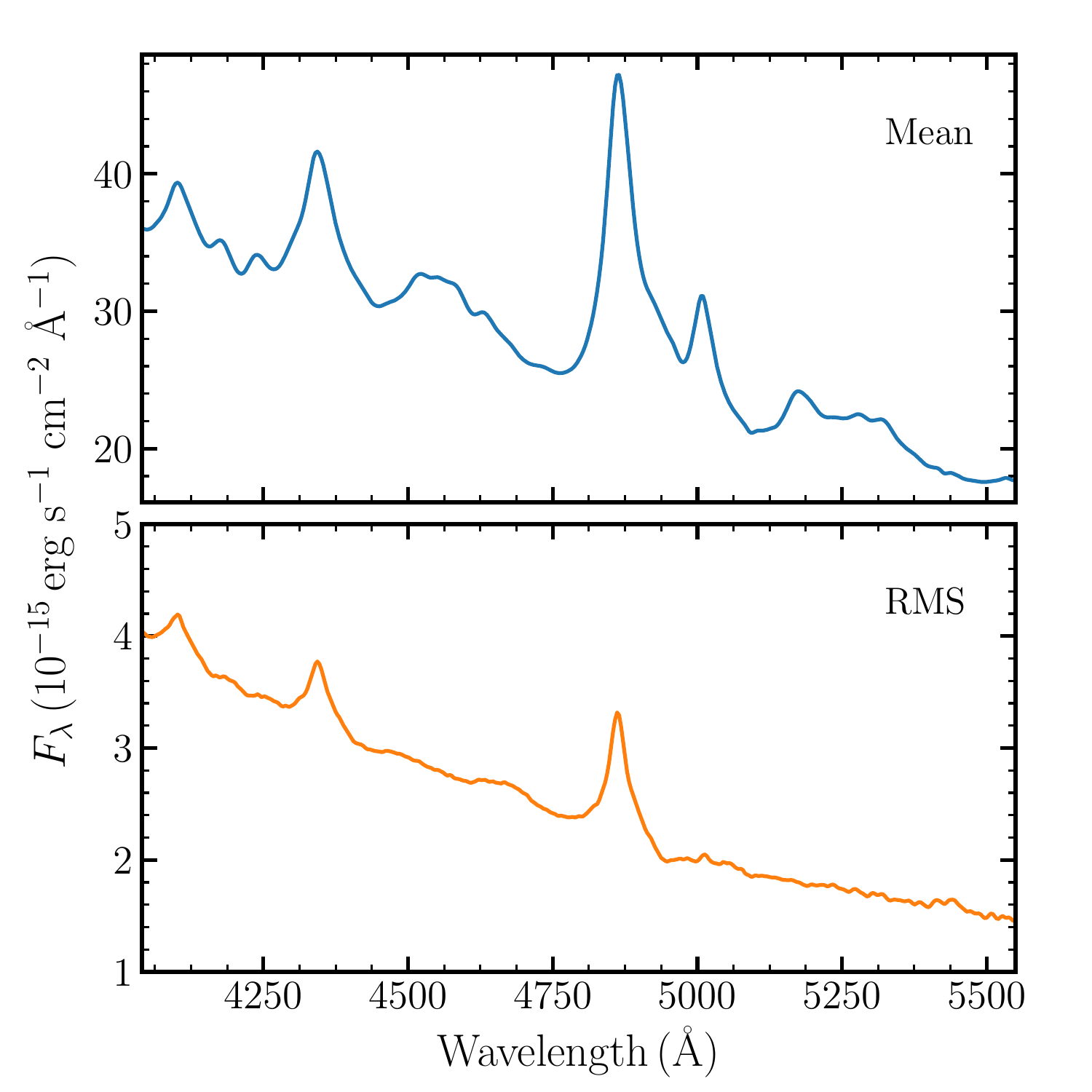}
	\caption{
		\footnotesize
		Mean and RMS spectra of 3C~273 (observed flux vs. rest-frame wavelength). \label{fig:meanrms}
	}
\end{figure}

To show the general properties of the spectra and evaluate the variations
at different velocities, we plot the mean and the root-mean-square
(RMS) spectra in Figure \ref{fig:meanrms}. 
The definitions of the mean and RMS spectra are
\begin{equation}
	\bar{F}_{\lambda}=\sum_{i=1}^NF_{\lambda}^i/N
\end{equation}
and
\begin{equation}
	S_{\lambda}=\left[\sum_{i=1}^N\left(F_{\lambda}^i-\bar{F}_{\lambda}\right)^2/N\right]^{1/2},
\end{equation}
respectively, with $N$ being the total number of spectra, 
and $F_{\lambda}^i$ being the $i$-th spectrum. The strong 
signatures of emission lines in the RMS spectrum indicate 
that their variations are significant. 

\subsection{Light curves}\label{sec:lightcurve}

\begin{figure}
	\centering
	\includegraphics[width=0.48\textwidth]{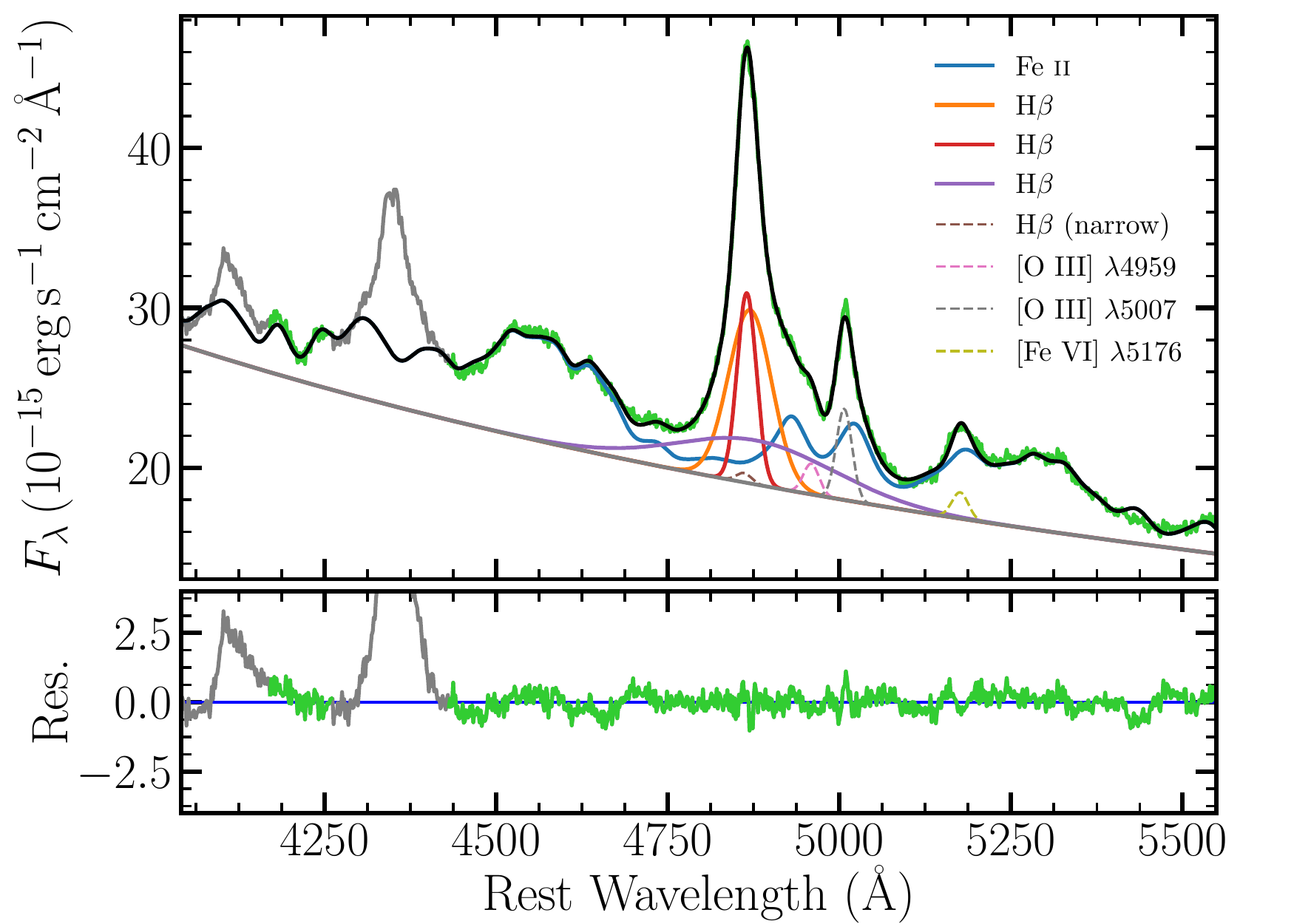}%fig_fit.pdf}
	\caption{
		\footnotesize
		An example of the spectral fitting. The part of the spectrum included
		in the fitting is plotted in green, excluded part is marked
		in gray. The best-fit model (in black) is composed of the power-law
		continuum, \feii\ emission (blue line, using the template from
		\citealt{Boroson1992}), broad \hbeta\ (red, orange, and purple lines),
		and a group of narrow emission lines (dashed lines). The residual is
		shown in the bottom panel.\label{fig:fit}}
\end{figure}

We use the multi-component spectral fitting approach to measure the light
curves of the emission lines \citep[e.g.,][]{Barth2013, Hu2015, Hu2016}. The
continuum is modeled with a power-law and the narrow emission-line components
(\oiiid\ and narrow \hbeta) by a single Gaussian.  The emission features of
\feii\ are modeled using the template from  \cite{Boroson1992}. The profile of
broad H$\beta$ line is modeled by 3 Gaussians. In addition, there are some
other narrow emission lines in the optical \feii\ region \citep{Vanden2001}.
The high-S/N  spectra of 3C~273 allow us to add extra line to  improve the
fitting, which can also compensate the potential inconsistency between the
\feii\ of 3C~273 and  the \feii\ template of I~Zw~1 from \cite{Boroson1992}.
Similar to \citet{Hu2015},  we add a coronal line [Fe {\sc vi}]$\lambda5176$
in our fitting. All of the narrow lines are fixed to have the same width and
shift. The \oiii$\lambda$4959 and the narrow H$\beta$ are fixed to have $1/3$
\citep{Oster2006} and $1/10$ \citep{Kewley2006,Stern2013} of the
\oiii$\lambda$5007 flux, respectively. We tried to fit the broad \hbeta\
using two Gaussians, but got poor fitting results with significant broad
\hbeta\ signal in the residual spectra,  which means two  Gaussians are not
enough to describe the \hbeta\ profile. Therefore, we changed to using three
Gaussians in the fitting.  The contribution of the host galaxy in the slit is
estimated to be $\lesssim6\%$ (estimated from our image fitting in Section
\ref{sec:stellarmass}), thus is  ignored in the fitting. The fitting is
performed mainly at 4430-5550\AA\ (the \hbeta\ and \feii\ region), and the
window 4170-4260\AA\ is also added  to constrain the continuum slope
\citep{Hu2008}. We do not add the \hgamma\ region in the fitting window in
order to reduce the number of model components.  An example of a fit is shown
in Figure \ref{fig:fit}.

After subtracting the continuum, \feii, \oiii$\lambda\lambda$4959,5007, and
[Fe {\sc vi}]$\lambda5176$,  the light curves of the H$\beta$ and H$\gamma$
emission lines  are measured by integrating the fluxes of the residual
spectra in the windows 4800-4920\AA\ and 4280-4420\AA\ in the  rest frame,
respectively. We do not subtract narrow \hbeta\ in the integration window;
thus, the \hbeta\ light curve contains the contribution from its narrow
component. Considering that the variability time scale of narrow line is much
longer,  it does not influence the lag measurement. The 5100\AA\ continuum and
the \feii\ (4434-4684\AA)  light curves are derived directly from the fitting
results.
	
\begin{figure*}[ht]
	\begin{center}
		\includegraphics[width=0.85\textwidth]{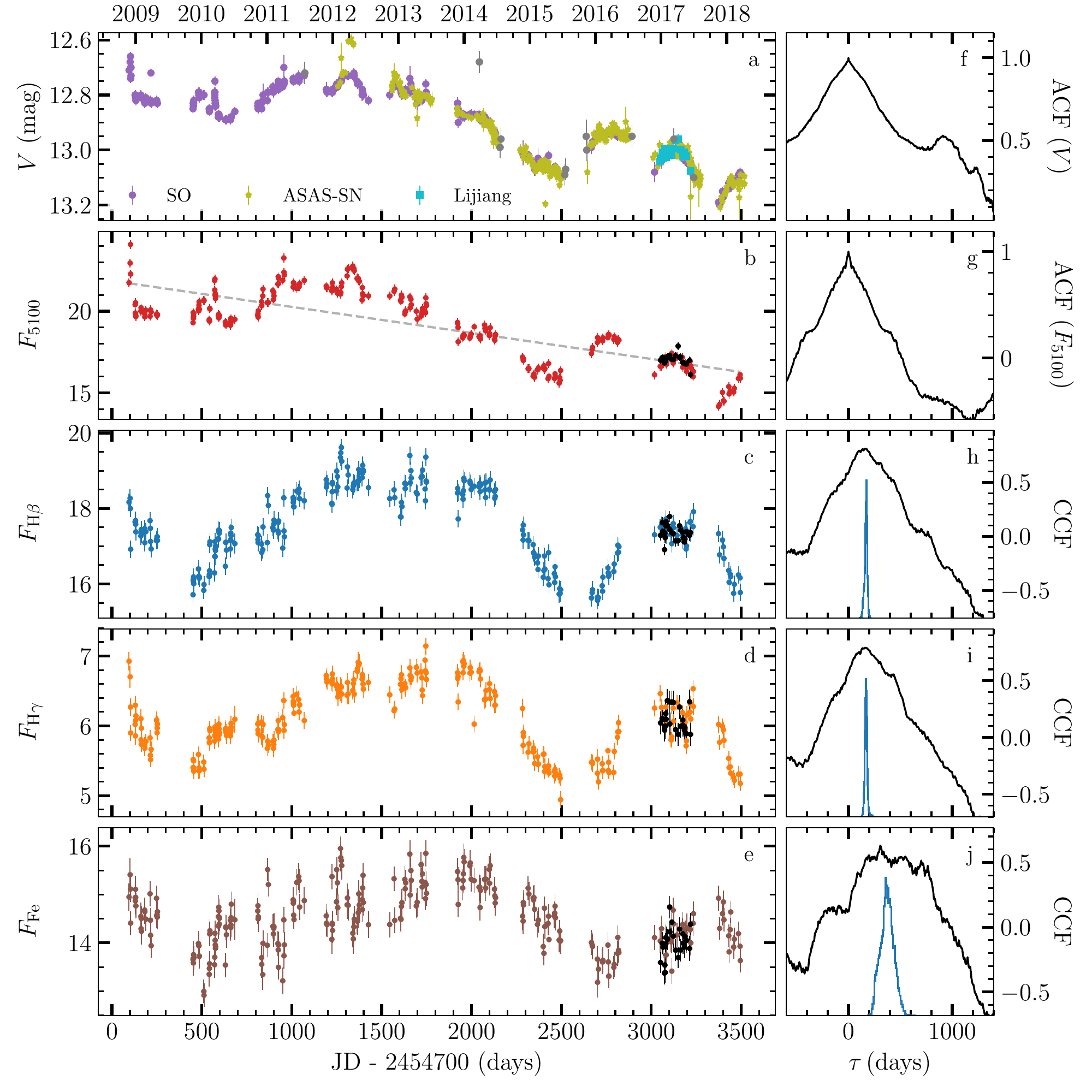}
	\end{center}
	\caption{
		\footnotesize
		Light curves and CCFs. The left panels are the light curves of the
		$V$-band, the continuum at 5100\AA, and the various emission lines.
		The $V$-band light curve was observed by SO, ASAS-SN, and Lijiang. 11
		points in gray are removed in the time-series analysis of this  paper
		(see more details in Section \ref{sec:lightcurve}). The $F_{5100}$
		light curve is in units of  ${\rm
		{10}^{-15}erg\,s^{-1}\,{cm}^{-2}\,\AA^{-1}}$, and emission-line fluxes
		are in  ${\rm {10}^{-13}erg\,s^{-1}\,{cm}^{-2}}$.  Here
		\feii\, is the blue blend from 4430\AA\ to 4680\AA. The gray dashed line
		in the 5100\AA\ light curve is the linear fit for detrending. The
		black points in panels b-f are obtained from Lijiang after the inter-calibration 
		(see  Section \ref{sec:intercal}). All of the other
		spectroscopic data points are from the SO observations.  In the right
		panels, black curves denote the CCFs (or ACFs).  Blue histograms show
		the corresponding CCCD. The CCFs are obtained from $F_{5100}$ (after
		detrending) and the corresponding emission-line light curves. 
		The January 1 of each year is marked on the time axis of the upper panel.
		\label{fig:lightcurve}}
\end{figure*}

Spectra obtained in very poor weather  conditions, or at extremely large
airmasses are removed from the analysis.  This includes 11 SO spectra and
three spectra from Lijiang. In addition,  some data points in the $V$-band
light curve were edited out of the  analysis for similar reasons and are
colored gray in Figure \ref{fig:lightcurve}. Error bars shown in the light
curves contain both Poisson noise and systematic uncertainties\footnote{In the
$V$-band light curve, we do not take into account the uncertainty  of the
magnitude of the photometric standard star (Star C). It only causes a
systematic shift by the same   amount for all of  the points in the light
curve.}. The systematic uncertainties are estimated by  the median filter
method \citep{Du2014}. We first smooth the light curve by a  median filter
with 5 points, and then subtract the smoothed light curve from the original
one. The standard deviation of the residual is used as the estimate of the
systematic uncertainty.

3C~273 emits strong radio and high-energy radiation
\citep[e.g.,][]{Cour1998,Turler1999,Soldi2008}. Fortunately, the contribution
from the synchrotron light in the optical band is weak
\citep{Impey1989,Smith1993}, which is also supported by the  non-blazar-like
flux variations in its  optical band. Meanwhile, the SO polarimetry shows that
its polarization (potentially produced by the synchrotron emission) remains
exceptionally low, except at the beginning of the program.

Furthermore, the RMS spectrum in Figure \ref{fig:meanrms} shows stronger
signal in the blue part, which means 3C~273 is more  variable at shorter
wavelengths. This is also evidence for the dominant contribution from the
accretion disk in the optical band, because the non-thermal emission from the
jet is much redder. The photometric monitoring campaign from  \citet{zeng2018}
also confirms this result. In addition, the variability  analysis in
\citet{Chidiac2016,Chidiac2017} did not  show any correlation between the
optical/IR and the radio/X-ray/$\gamma$-ray  emission, which also suggests the
optical flux is dominated by the thermal emission  from the accretion disk.
Only a few flare-like events are found in the $V$-band light curve (e.g.,
Julian date 100 and 580, from the zero point of 2454700 in Figure
\ref{fig:lightcurve}), but they do not cause serious disruption to the time-series 
analysis. Therefore, the jet emission of 3C~273 does not influence the
lag measurements in the following sections.

\subsection{Inter-calibration}\label{sec:intercal}

Because of the different apertures used for the SO, Lijiang, and ASAS-SN
observations,  inter-calibration of the three light curves is necessary. We
adopt the method used by \cite{Peterson1991} to scale the Lijiang light curve
to match the SO light  curve. For the continuum, we assume
\begin{equation}
	F_\mathrm{S}(\mathrm{5100 \mbox{\AA}}) = \varphi_{5100} F_\mathrm{L}(\mathrm{5100 \mbox{\AA}})-G_0,
\end{equation}
where $G_0$ is a constant to account for the difference in the contribution of
the host galaxy (aperture effect),  $\varphi_{5100}$ is a scale factor,
$F_\mathrm{S}(\mathrm{5100 \mbox{\AA}})$ and  $F_\mathrm{L}(\mathrm{5100
\mbox{\AA}})$ are the 5100\AA\ continuum fluxes of adjacent epochs ($<$ 4 day
apart) of the SO and Lijiang observations. For the Balmer lines and \feii\
emission, we assume
\begin{equation}
	F_{\rm S}({\rm line}) = \varphi_{\rm line} F_{\rm L}({\rm line}), \label{eq2}
\end{equation}
where ``line'' refers to the relevant emission line, $\varphi_{\rm line}$ is
the scale factor, and $F_{\rm S}$ and $F_{\rm L}$ are the emission-line fluxes
measured for epochs $<$ 4 days apart. Considering that the central wavelength
of \hbeta\ is not far from  5100\AA, we assume 
$\varphi_{\rm H\beta}=\varphi_{5100}$. Using the Levenberg-Marquardt algorithm
\citep{Press1992}, the best-fit values are  $\varphi_{5100, \rm H\beta} =
0.9986$, $\varphi_{\rm H\gamma}=1.0571$, $\varphi_{\rm FeII}=0.9906$, and 
$G_0={\rm 0.02\times10^{-15}}$ ${\rm erg\ s^{-1}\ cm^{-2}\ \mbox{\AA}^{-1}}$. The
small value of $G_0$ indicates that the aperture effect with respect to  the
host galaxy contribution to the spectrum is negligible and that the overall
contamination by the host galaxy to the data is weak  (see Section
\ref{sec:SOPS}). The calibrated light curves are shown in Figure
\ref{fig:lightcurve} and provided in Table \ref{table:lightcurve}. Similarly,
the ASAS-SN and Lijiang $V$-band light curves shown in Figure
\ref{fig:lightcurve} have also been scaled to match the SO $V$-band light
curve.
	
\section{Analysis}\label{sec:res}

\subsection{Variability characteristics}
The $F_{\rm var}$ parameter \citep{Rodr1997,edelson2002} is adopted
to quantify the variability amplitude of the light curve. It is defined as
\begin{equation}
	F_{\rm var} = \frac{\left(\sigma^2-\Delta^2\right)^{1/2}}{\langle f\rangle} ,
\end{equation}
where $\langle f\rangle$ is the mean flux in the light curve,
$\sigma$ is the square root of the variance
\begin{equation}
	\sigma^2 = \frac{1}{N-1}\sum_{i=1}^{N}\left(f_i - \langle f\rangle\right)^2 ,
\end{equation}
and $\Delta^2$ is the mean square value of the uncertainties $\Delta_i$
\begin{equation}
	\Delta^2 = \frac{1}{N}\sum_{i=1}^{N}\Delta_i^2 .
\end{equation}
The $F_{\rm var}$ of each light curve is listed in Table \ref{table:stas}.
The variation amplitudes of the continuum and \hbeta\ fluxes are
10.9\% and 5.2\%, respectively. 
In addition, we also list $R_{\rm max}$ in 
Table \ref{table:stas}, which is defined as the ratio of the maximum to the
minimum in the light curve, as well as the
median values of the light curves and their standard deviations.

\begin{deluxetable}{lrrc}
\tabletypesize{\footnotesize}
	\tablecaption{Light Curve Statistics \label{table:stas}}
	\tablecolumns{4}
	%\tabletypesize{\scriptsize}
	\tablewidth{0pt}
%	{\footnotesize
		\tablehead{
			\colhead{Lines} &
			\colhead{Median Flux} &
			\colhead{$F_{\rm var}(\%)$} &
			\colhead{$R_{\rm max}$}
		}
		\startdata
		\hbeta     & $17.40\pm0.94$ &  $5.24\pm0.23$ & 1.23 \\
		\hgamma    &  $6.02\pm0.47$ &  $7.44\pm0.33$ & 1.32 \\
		\feii      & $14.44\pm0.61$ &  $3.81\pm0.19$ & 1.18 \\
		$F_{5100}$ & $19.46\pm2.09$ & $10.88\pm0.45$ & 1.51 \\
		\enddata
		\tablecomments{\footnotesize
		The emission-line fluxes and
			5100\AA\, continuum are in units of
			${\rm 10^{-13}\,erg\,s^{-1}\,cm^{-2}}$ and
			${\rm 10^{-15}\,erg\,s^{-1}\,cm^{-2}\,\AA^{-1}}$, respectively. 
			The flux,
			$F_{\rm var}$, and $R_{\rm max}$ of the $F_{5100}$ light
			curve are measured in the observed frame (with Galactic extinction correction).}
\end{deluxetable}

\subsection{Time lags\label{section:timelag}}
	
The continuum light curve of 3C~273 shows long-term variations that do not
appear in the  emission-line light curves (see Figure \ref{fig:lightcurve}).
This long-term trend may bias the lag detection,  and thus we ``detrend'' the
5100\AA\ light curve by subtracting a linear fit before the lag measurement
(suggested by \citealt{Welsh1999}, see also, e.g., \citealt{Peterson2004,
Denney2010}). This fit is shown  in Figure \ref{fig:lightcurve}. The
detrending improves the correlation coefficients in the following cross-correlation 
analysis. We tried to detrend the emission-line light curves, but
the slopes of their  linear fits are all close to zero within 3-$\sigma$
uncertainties. Therefore, we only detrend the 5100\AA\ light curve.

We adopt the interpolated cross-correlation function \citep[ICCF; see][]
{gas1986, gask1987, white1994} to measure the time lags of the emission 
lines relative to the variation of the continuum at 5100\AA\ after detrending. The centroid of the CCF above 
a typical value (80\%) of the maximum correlation coefficient ($r_{\rm max}$) 
is used as the lag measurement.
			
The uncertainties of the time lags are obtained through the ``flux
randomization/random subset sampling (FR/RSS)'' method
\citep{Maoz1990,Peterson1998b,Peterson2004}, which both 
randomizes the measured fluxes based on their uncertainties and 
resamples the light curves.
The resampling is performed by randomly selecting $N$ points 
from the light curve (redundant selection is allowed), 
where $N$ is the total number of epochs in the light curve. A subsample is constructed
after ignoring the redundant points (roughly a fraction of $\sim0.37$). We repeat this process
10000 times and obtain the cross-correlation centroid 
distribution (CCCD) by performing CCF to the light-curve subsamples. 
It is similar to the so-called ``bootstrapping'' technique.
Then the values of 
the time lags and the corresponding uncertainties are
obtained from the median, 15.87\%, and 84.13\% ($1\sigma$) quantiles
of the CCCD.

The auto-correlation functions (ACFs), CCFs, and  CCCDs between the emission-line 
light curves and the detrended 5100\AA\ light curve  are shown in the
right panels of Figure \ref{fig:lightcurve}. We also provide the results
without  the detrending in Appendix. The time  lags of the emission lines are
listed in  Table \ref{table:timelag}. The time lag of the \hbeta\ emission
line is $146.8^{+8.3}_{-12.1}$ days in the rest frame,  consistent with the
lag found for the \hgamma\ line. The response of \feii\ to continuum
variations is found to be nearly a factor  of 2 longer than that of the Balmer
lines, at $322.0^{+55.5}_{-57.9}$ days in the rest frame. The lag
uncertainties in our measurements are smaller than those in \cite{Kaspi2000},
because of our denser sampling cadence, higher S/N ratios, and longer
monitoring period.

\begin{figure}
	\centering
	\includegraphics[width=0.468\textwidth]{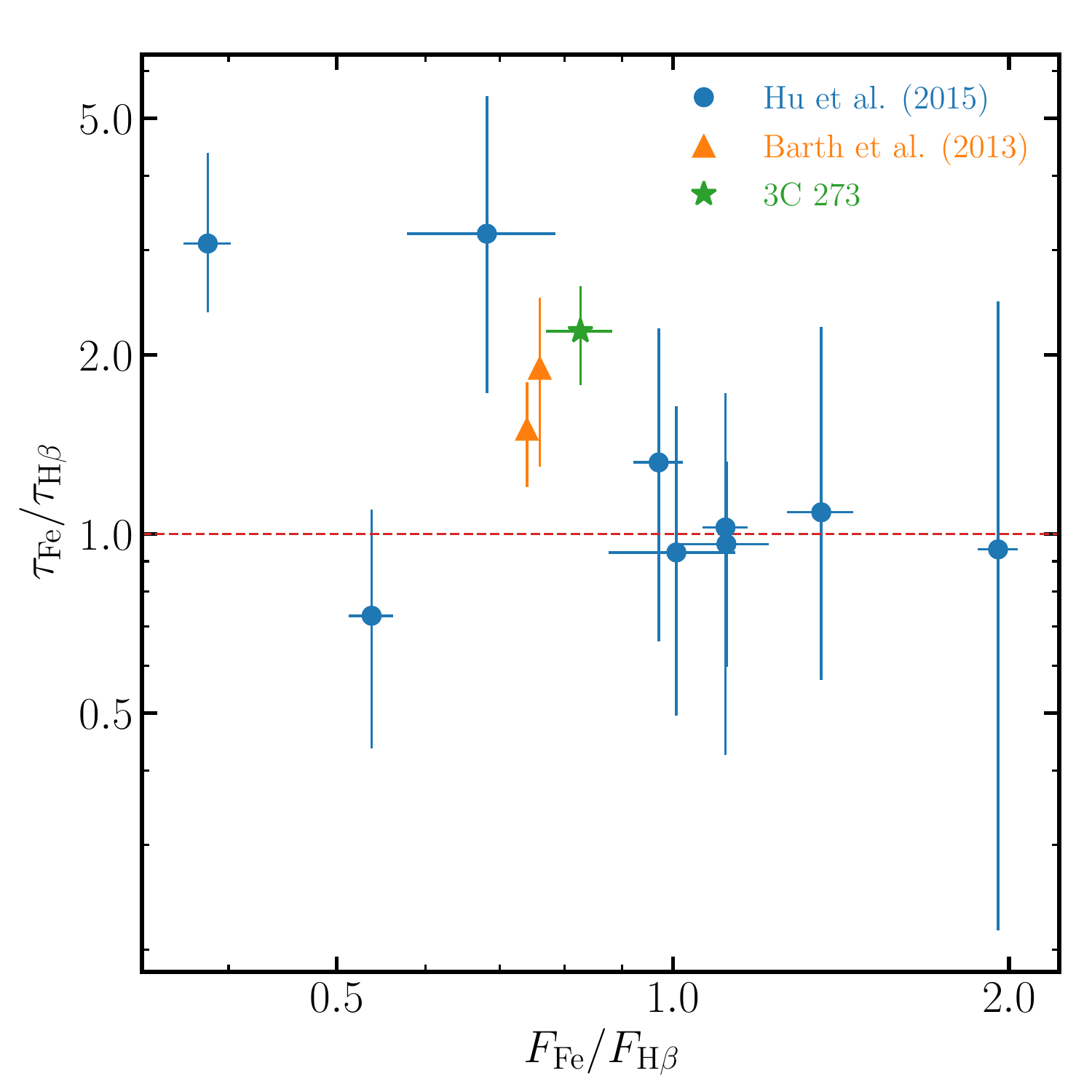}%{fig_feiibeta_detrend.pdf}
	\caption{\footnotesize
		Time lag ratio of \feii\ to \hbeta\ vs. their intensity ratio.
		The orange triangles show the two objects
		in \cite{Barth2013}. The blue circles are the
		objects presented in \cite{Hu2015}.}\label{fig:feiibeta}
\end{figure}
			
\citet{Hu2015} show a tentative correlation between the time 
lag ratio ($\tau_{\rm Fe}/\tau_{\rm H\beta}$) and the intensity ratio 
($F_{\rm Fe}/F_{\rm H\beta}$) of the \feii\ and \hbeta\ lines. The \feii\ lag is
similar to the \hbeta\ lag if $F_{\rm Fe}/F_{\rm H\beta}\gtrsim1$, while the
\feii\ lag tends to be longer if $F_{\rm Fe}/F_{\rm H\beta}<1$.
The \feii\ and \hbeta\ time lags of 3C~273 are in good agreement with
the $\tau_{\rm Fe}/\tau_{\rm H\beta}$ -- $F_{\rm Fe}/F_{\rm H\beta}$
correlation (Figure \ref{fig:feiibeta}).

\subsection{Velocity-resolved lags of \hbeta}\label{sec:velres}
			
We use velocity-resolved RM to investigate the geometry and kinematics of the 
BLR in 3C~273.
The \hbeta\ line is divided into 15 velocity bins, with each bin having 
the same flux as in the RMS spectrum 
created after the subtraction of continuum, \feii, and narrow lines in Section \ref{sec:lightcurve}
\citep[e.g.,][]{Bentz2008, Denney2010, Grier2013, Du2016VI, De2018}.
The fluxes in each bin are integrated to generate light curves that are used 
to cross-correlate with the 5100\AA\ variations (after detrending). The time lag and the 
associated uncertainties of each bin are obtained using the same method 
described in Section \ref{section:timelag}.

\begin{figure}
	\centering
	\includegraphics[width=0.46\textwidth]{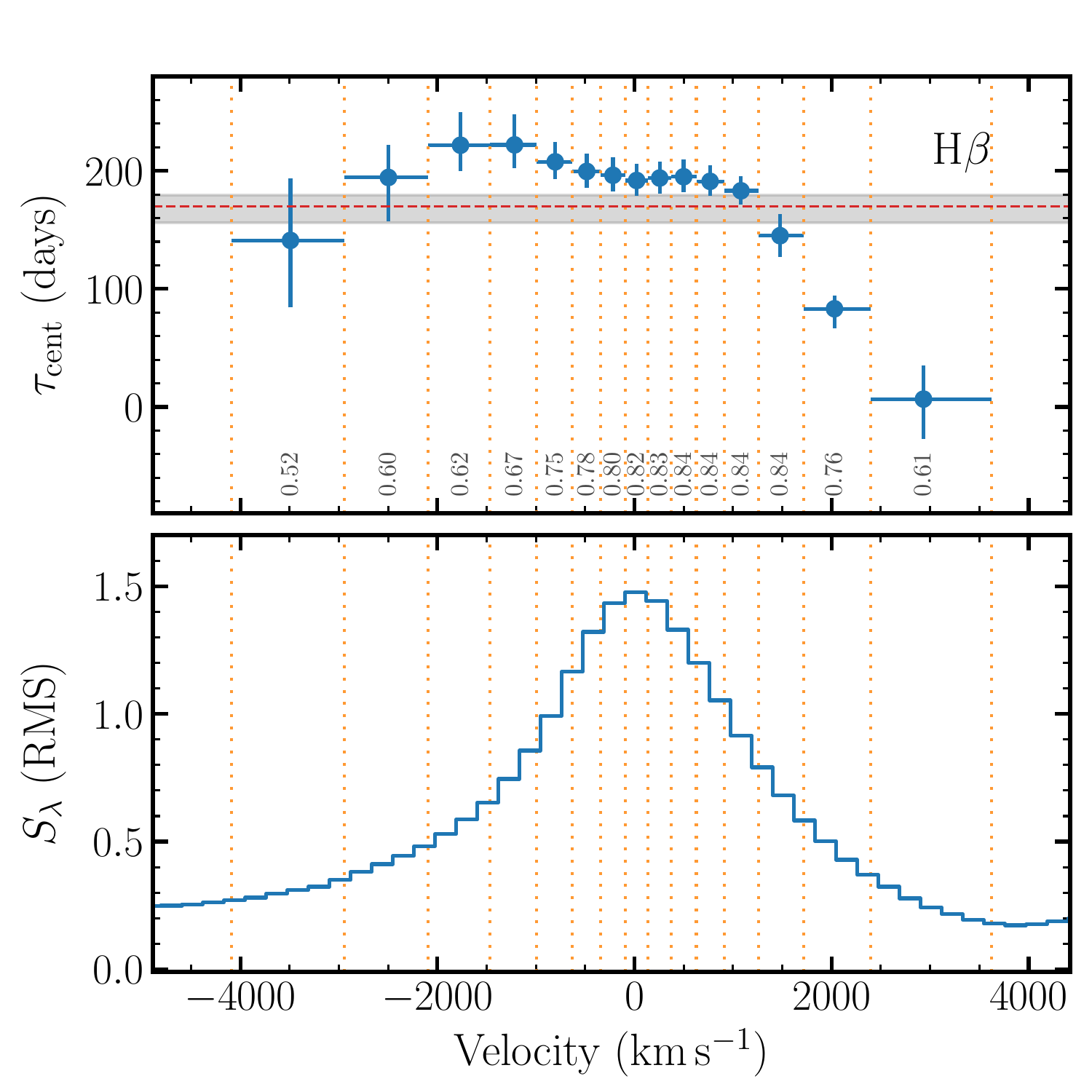}%{fig_velo_beta_detrend_f5100.pdf}
	\caption{\footnotesize
		Velocity-resolved time lags of the \hbeta\ line
		in the observed frame. The top panel shows the
		centroid time lags and their uncertainties in velocity
		bins denoted by the vertical dotted lines. 
		The horizontal dashed line and the shadowed region
		are the mean time lag and its associated 1-$\sigma$ 
		uncertainties. 
		The gray numbers in the top panel are the maximum correlation coefficients 
		of the CCFs in the velocity bins.
		The RMS spectrum is shown in the bottom
		panel in units of 
		${\rm 10^{-15}\,erg\,s^{-1}\,cm^{-2}\,\AA^{-1}}$.
	}\label{fig:velo}
\end{figure}

We plot the velocity-resolved time lags (in the observed frame) 
in Figure \ref{fig:velo}. The maximum correlation coefficients for individual bins are also 
shown in Figure \ref{fig:velo}. The velocity-resolved result shows a complex structure.
In general, the lags are longer at small velocities and shorter at high velocities, which
is the signature of a Keplerian disk or virialized motion. However, the lags at the blue velocities are
systematically longer than those in the red wing. There are at least several possibilities:
(1) a classic explanation is that the \hbeta-emitting gas has some inflowing radial 
velocity \citep[e.g.,][]{Bentz2008, Denney2010, Grier2013, Du2016VI, De2018}; (2) a
rotating disk wind model suggested by \cite{Mangham2017} can produce
the velocity-resolved lags resembling the ``red-leads-blue'' inflow signature; and
(3) the BLR gas is located in an eccentric or lopsided disk, and shows net velocity along the
line of sight. The BLR dynamical modeling technique \cite[e.g.,][]{Pancoast2011, Pancoast2012, 
Grier2017, Williams2018} is required to investigate the geometry and kinematics of the BLR of 3C~273 
in more detail in the future. 

\begin{deluxetable}{lccc}
\tabletypesize{\footnotesize}
	\tablecaption{Emission-line Time Lags\label{table:timelag}}
	\tablehead{
\footnotesize
		\multirow{2}{*}{Lines} &
		\multirow{2}{*}{$r_{\rm max}$} &
		\multicolumn{2}{c}{lags (days)}\\ \cline{3-4}
		\colhead{} &
		\colhead{} &
		\colhead{observed frame} &
		\colhead{rest frame}
	}
	\startdata
	\hbeta  & 0.81 & $170.0_{-14.0}^{+9.6}$ & $146.8_{-12.1}^{+8.3}$ \\
	\hgamma & 0.79 & $169.7_{-11.2}^{+10.2}$ & $146.5_{-9.7}^{+8.8}$ \\
	\feii   & 0.63 & $373.0_{-67.1}^{+64.3}$ & $322.0_{-57.9}^{+55.5}$ \\
	\enddata
\end{deluxetable}
				
\subsection{Line widths}\label{sec:linewidth}

We investigate the behavior of the broad emission-line profiles of \hbeta\ 
and \hgamma\ through the measurement of the FWHM 
and the line dispersion ($\sigma_{\rm line}$). The line dispersion is defined as
\begin{equation}
	\sigma_{\rm line}^2 (\lambda) = \left<\lambda^2\right> - \lambda_0^2,
\end{equation}
where $\lambda_0 = \int \lambda f(\lambda)d \lambda / \int f(\lambda)d\lambda$ 
is the flux-weighted centroid wavelength of the emission line profile,
$\left< \lambda^2\right> = \int \lambda^2 f(\lambda) d\lambda / \int f(\lambda) d\lambda$,
and $f(\lambda)$ is the flux density. Line widths are measured 
from the line profiles in the mean and RMS spectra after 
subtraction of the continuum, \feii\ emission, and narrow emission lines.
The uncertainties of the line widths are estimated by the
bootstrap method \citep{Peterson2004}. We
randomly select $N$ spectra with replacement (redundant selection is 
allowed, $N$ is the total number of spectra
for 3C~273), and construct a spectral subset
after removing the redundant spectra.
Then, the mean and RMS spectra are generated from the subset,
and are used to measure the line widths. This process is repeated
10000 times to create the line width distributions.
The standard deviations of the FWHM and $\sigma_{\rm line}$
distributions are regarded as the uncertainties.
				
The line widths of \hbeta\ and \hgamma\ are calculated
separately for the SO and Lijiang data.
To estimate the spectral broadening caused by the instruments and seeing, 
we compare our spectra with higher-resolution spectra obtained with an 
instrument with a well-understood line spread function.  In this case, we use 
the {\tt o44301010} and {\tt o44301020} observations of 3C~273 made by STIS 
aboard the Hubble Space Telescope \citep{Hutchings1998}.
The spectral broadening of SO and Lijiang are
$1157\pm46$ km/s and $485\pm38$ km/s, respectively. After subtracting
the spectral broadening in quadrature, the line widths of the
\hbeta\ and \hgamma\ lines are listed in Table
\ref{table:linewidth}. For the SO spectra, the FWHM of 
\hbeta\ and \hgamma\ are both
around $\sim$3300 km/s. We adopt the FWHM from the mean spectrum to
calculate of BH mass in the following sections,
because the $\sigma_{\rm line}$ measurement is sensitive to the wings of
the emission line and the accuracy of the continuum subtraction. 
The FWHM of \feii\ is 2142.4$\pm$116.1 km/s, 
which is obtained directly from the fitting. Compared with the \hbeta\ 
and \hgamma, the FWHM of \feii\ is significantly narrower. This suggests that 
the \feii-emitting region is farther away from the BH than the region 
in the BLR producing the Balmer emission. 

\begin{deluxetable*}{llccccc}
\tabletypesize{\footnotesize}
	\centering
	\tablecaption{Emission-line Widths \label{table:linewidth}}
	\tablehead{
		\multirow{2}*{} &
		\multirow{2}*{Lines} &
		\multicolumn{2}{c}{Line width (mean)} &
		\colhead{} &
		\multicolumn{2}{c}{Line width (RMS)} \\
		\cline{3-4}
		\cline{6-7}
		\colhead{} &
		\colhead{} &
		\colhead{FWHM (${\rm km\,s^{-1}}$)} &
		\colhead{$\sigma_{\rm line}$ (${\rm km\,s^{-1}}$)} &
		\colhead{} &
		\colhead{FWHM (${\rm km\,s^{-1}}$)} &
		\colhead{$\sigma_{\rm line}$ (${\rm km\,s^{-1}}$)}
	}
	\startdata
	\multirow{3}*{Steward} & \hbeta & $3314.1\pm59.3$ & $1698.8\pm25.1$ & & $1941.4\pm69.5$ & $1098.9\pm39.9$ \\
	~ & \hgamma & $3313.8\pm59.2$ & $1667.7\pm25.2$ & & $2439.1\pm82.5$ & $1444.5\pm46.4$\\
	~ & \feii & $2142.4\pm116.1$ & $\cdots$ & & $\cdots$ & $\cdots$\\
	\cline{1-7}
	\multirow{3}*{Lijiang} & \hbeta & $3196.9\pm39.2$ & $1702.9\pm16.5$ & & $3305.2\pm420.1$ & $1081.0\pm238.8$ \\
	~ & \hgamma & $3139.0\pm40.5$ & $1535.9\pm19.6$	& & $4079.7\pm509.2$ & $1176.9\pm260.8$\\
	~ & \feii & $2039.3\pm70.6$ & $\cdots$ & & $\cdots$ & $\cdots$\\
	\enddata
\end{deluxetable*}

\begin{figure}
	\centering
	\includegraphics[width=0.46\textwidth]{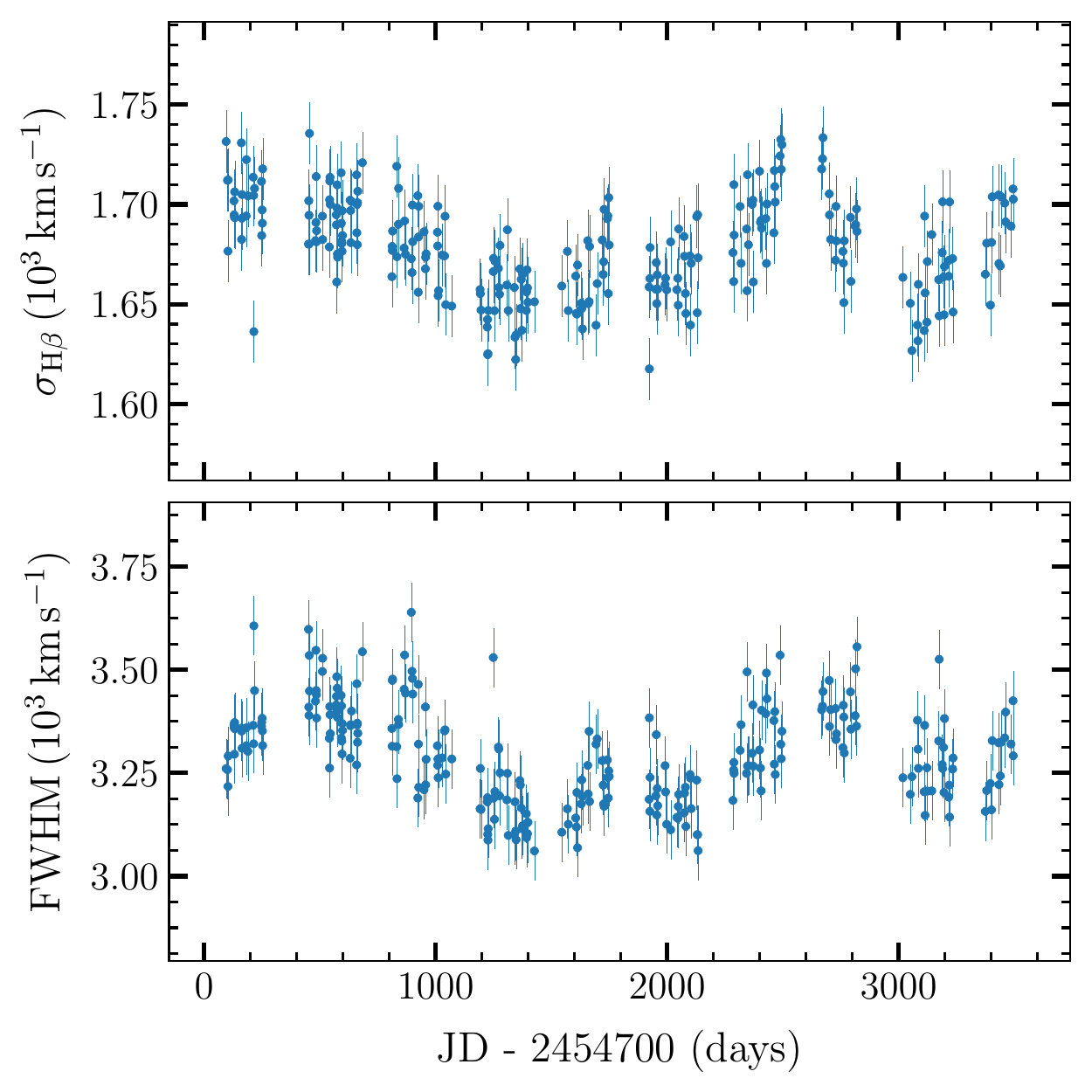}%{fig_jd_fwhm_sigma_beta.pdf}
	\caption{\footnotesize
		Variability of the \hbeta\ profile in
		3C~273. Two measures of the \hbeta\ width from SO are plotted with
		respect to time over a period of almost 10 years. 
		The error bars of the line widths are estimated by the median filter method
		(see more details in Section \ref{sec:lightcurve}).
		\label{fig:jd_fwhm}}
\end{figure}

\begin{figure}
	\centering
	\includegraphics[width=0.46\textwidth]{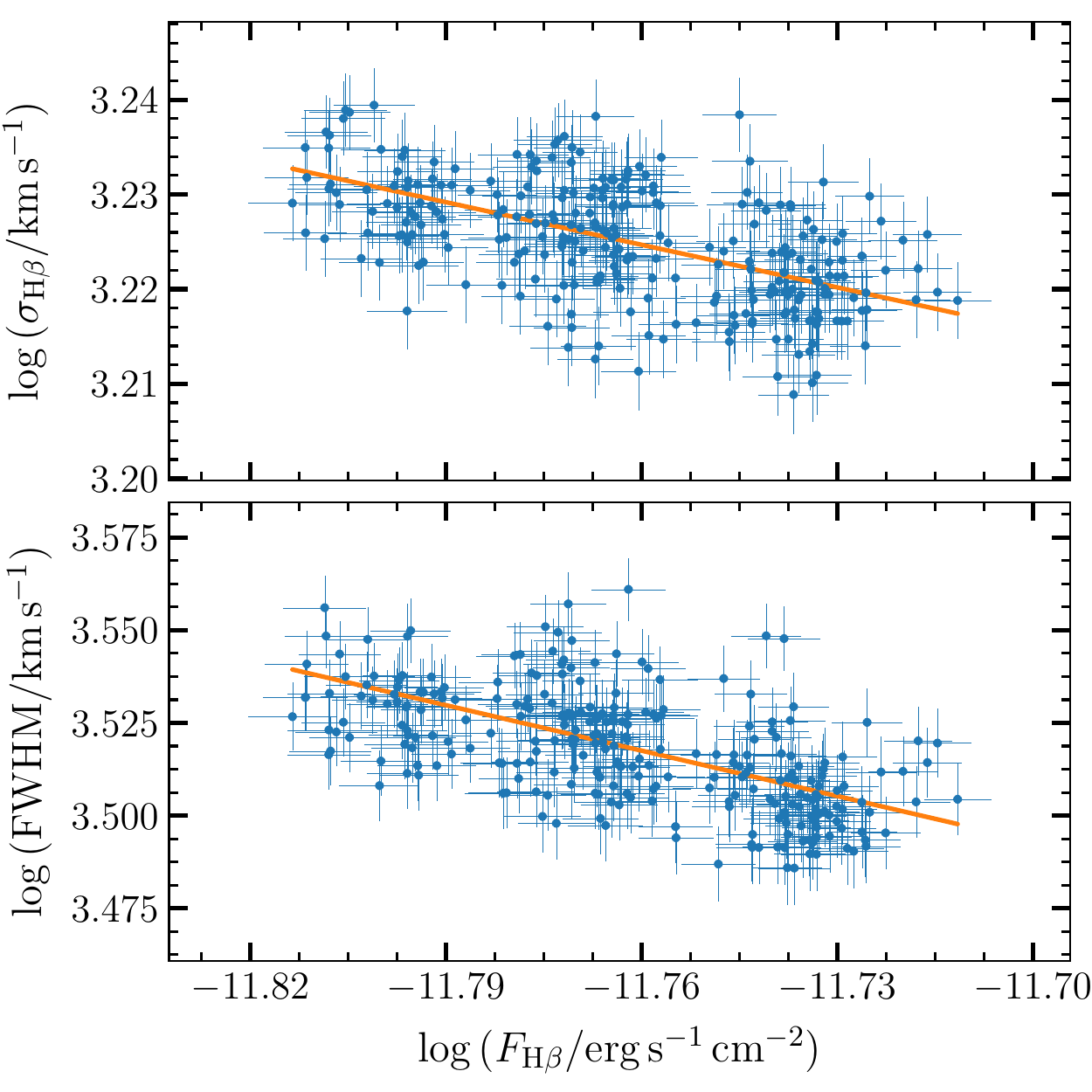}%{fig_flux_fwhm_sigma_beta.pdf}
	\caption{\footnotesize
		Emission-line width of \hbeta\ as a function of its flux (for the SO program).
		The orange lines are the best fits to the data. 
		The error bars of the line widths are same as in Figure \ref{fig:jd_fwhm}.
		\label{fig:flux_linewidth}
	}
\end{figure}
				
The FWHM and $\sigma_{\rm line}$ of individual spectra are also measured to
investigate the line width changes during the campaign. After correcting for
spectral broadening, the FWHM and $\sigma_{\rm line}$ of \hbeta\ from SO for the
entire monitoring period are shown in Figure \ref{fig:jd_fwhm}. Figure
\ref{fig:flux_linewidth} plots the correlation between \hbeta\ emission-line
flux and line width. There is a strong anti-correlation between  line flux and
width in \hbeta\ (Pearson's coefficient and the null probability  are $-0.59$
and $3.8 \times 10^{-27}$ for FWHM vs. $F_{\rm H\beta}$,  and are $-0.54$ and
$4.5 \times 10^{-22}$ for $\sigma_{\rm line}$ vs. $F_{\rm H\beta}$,
respectively.) To quantify the correlation, we perform the linear regression
of
\begin{equation}
	\log V_{\rm H\beta} = \alpha + \beta \log F_{\rm H\beta},
	\label{eq:flux_linewidth}
\end{equation}
where $V_{\rm H\beta}$ is $\rm FWHM_{H\beta}$ or $\sigma_{\rm H\beta}$.
The uncertainties of the parameters are obtained using the bootstrap technique.
The results are shown in Figure \ref{fig:flux_linewidth}.
The best-fit values for ($\alpha$, $\beta$) are ($-1.28 \pm 0.37$, $-0.41 \pm 0.03$)
and ($1.46 \pm 0.15$, $-0.15 \pm 0.01$) for
flux versus FWHM and $\sigma_{\rm H\beta}$, respectively.
Similar correlations are reported for some other objects
\citep[e.g.,][]{Barth2015}.

\subsection{Line width versus lags}\label{sec:lwvslags}
				
\begin{figure}
	\centering
	\includegraphics[width=0.46\textwidth]{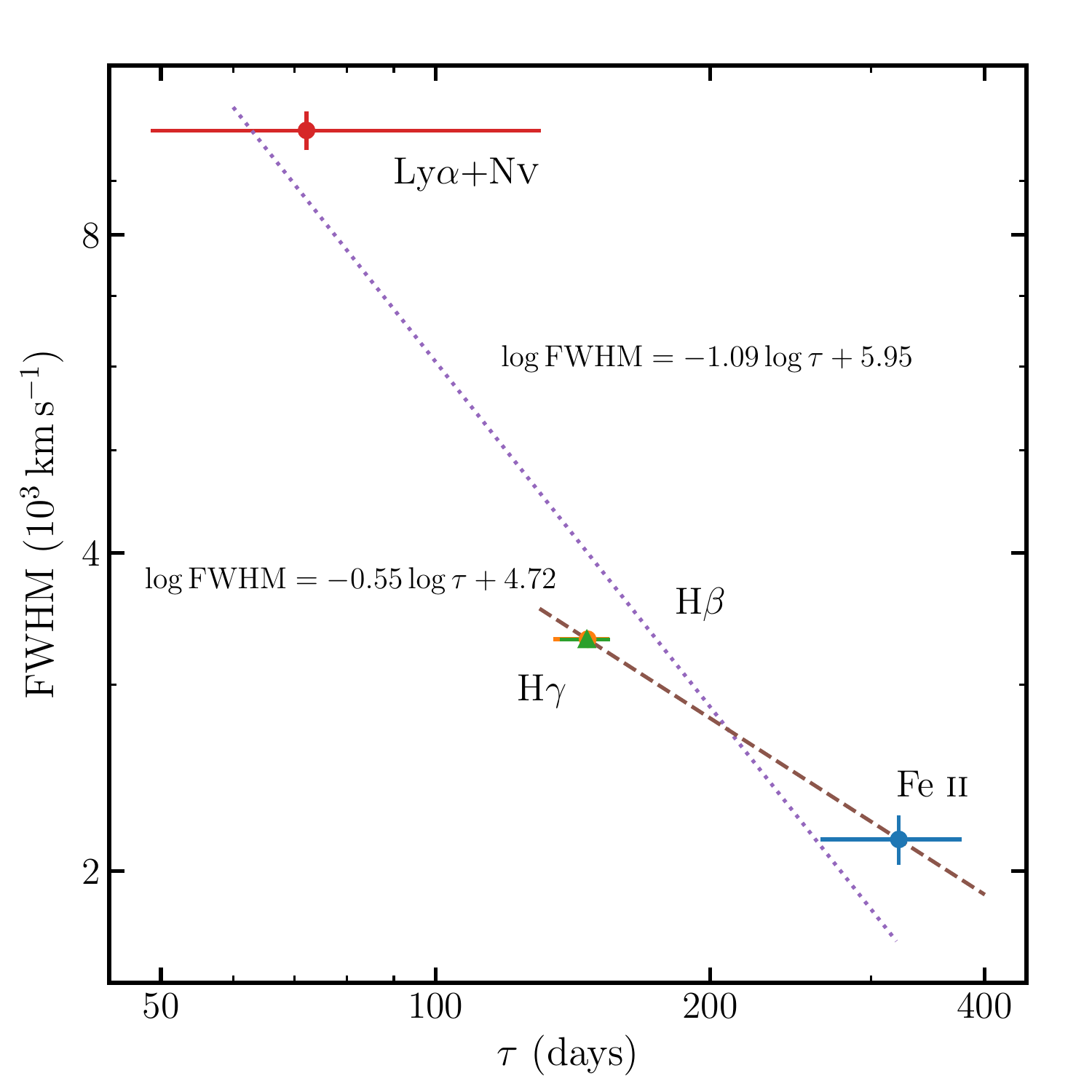}%{fig_dynamical_detrend.pdf}
	\caption{\footnotesize
		Emission-line widths vs. their rest-frame time lags. 
		The FWHM of \lya+ \Nv\ is taken from
		\citet{Paltani2005}, with the time lag recalculated in
		Section \ref{sec:lwvslags}. The best linear fit for all emission lines 
		is shown by the dotted line. The dashed line represents the best fit 
		that omits \lya+ \Nv.
		}\label{fig:fwhmtau}
\end{figure}

The RM campaign of 3C~273 performed by the {\it International
Ultraviolet Explorer} ({\it IUE}) yielded time lags
of $\tau_{\rm \lya+\Nv} = 435_{-77}^{+77}$ and
$\tau_\civ = 690_{-106}^{+102}$ days in the rest frame for \lya+ \Nv\ and 
\civ lines relative to the UV continuum variation
(\citealt{Paltani2005}). The echo of the Ly$\alpha$ + \Nv\ lines with respect to
the UV continuum is quite good with $r_{\rm max}\approx 0.6$, but the
\civ response remains largely uncertain. \cite{Paltani2005} took the
centroid lag above $0.6\,r_{\rm max}$ of the \lya+ \Nv\ CCF as the lag
measurement. For comparison to our optical measurements, we re-calculated
their \lya+ \Nv\ lag using the $0.8\,r_\mathrm{max}$ criterion.
There are multiple peaks in the ultraviolet CCF, and we only adopt
the highest peak in the calculation.
The re-determined \lya+ \Nv\ lag is $\tau_{\rm Ly\alpha+ \Nv} = 72.2_{-23.5}^{+58.1}$ 
days in the rest frame. We plot the correlation between the lags and the 
FWHMs both for the UV and optical lines in Figure \ref{fig:fwhmtau}
, where it is seen that ${\rm FWHM}\propto \tau^{-0.55\pm 0.01}$ for 
the optical emission lines. This is similar to the results for NGC~5548 \citep{Peterson1999}. This
relation steepens to ${\rm FWHM}\propto \tau^{-1.1\pm 0.3}$ with the 
inclusion of the \lya+ \Nv\ lines.

3C~273 shows a ${\rm FWHM}\propto\tau^{-0.56}$ relation for its optical
emission lines that the UV emission lines do not follow. It implies that the
geometry or kinematics of the UV-emitting clouds are different from the
optical-line region, i.e., their virial factors in the BH mass measurements
should be very different. The optical emission lines are  relatively closer to
${\rm FWHM}\propto \tau^{-0.5}$ relation (consistent with  the velocity-resolved 
result that the  kinematics in the \hbeta\ region is generally a
Keplerian disk or virialized motion);  thus, we prefer to estimate the BH mass
based on those lines.  Given that the {\it IUE} campaign was carried out more
than a decade before the optical RM program, the steeper slope may also be
caused by a physical change in the geometry and/or kinematics of the BLR.

\section{Black hole mass and accretion rate}
\label{sec:bhmass}
%\subsection{Mass}
Combining the results for the time lag, $\tau_{\rm line}$, and the line width,
$V_{\rm line}$, the BH mass can be estimated by 
\begin{equation}\label{bhm}
	\bhm = f_{\rm BLR}\frac{ R_{\rm BLR}V_{\rm line}^2}{G},
\end{equation}
where $R_{\rm BLR}=c\tau_{\rm line}$ is the emissivity-weighted radius of
the BLR, $V_{\rm line}$ is given by the FWHM or $\sigma_{\rm line}$, 
$c$ is the speed of light, $G$ is the gravitational constant, 
and $f_{\rm BLR}$ is the virial factor depending on the kinematics, geometry, 
and inclination angle of the BLR.
In general, the average $f_{\rm BLR}$ can be estimated
by comparing RM-based BH masses for AGNs with available bulge stellar velocity 
dispersions with BH masses predicted from the $M-\sigma$ relation of inactive galaxies
\citep[e.g.,][]{Onken2004,Ho2014,Woo2015,Park2012,Grier2013,Ho2014}.
Specifically, for FWHM of \hbeta\ measured from their mean spectra, \citet{Ho2014} 
demonstrate that the $f_{\rm BLR}$ is $1.3$ for AGNs with classical bulges or elliptical 
host galaxies (see also \citealt{Mejia2018}).

Independent from the calibration based on the $\bhm-\sigma_{*}$ relation,
the Bayesian-based BLR dynamical modeling method
\citep{Pancoast2011,Pancoast2012} has been adopted in recent years to provide $f_{\rm BLR}$ for
individual AGNs with different BLR geometry and kinematics
\citep[e.g.,][]{Pancoast2014a,Pancoast2014b,Pancoast2018,Grier2017,Li2013,Li2018,Williams2018}.
The velocity-resolved time lags 
of 3C~273 show ``red-leads-blue'' signatures, which may arise from inflow kinematics
(in Section \ref{sec:velres}).  Virial factors for AGNs with inflowing signatures as derived from
dynamical modeling are diverse and range from $0.58-3.98$ \citep{Grier2017,
Pancoast2014b}, but they average close to the value found by
\citet{Ho2014}. Thus, we simply adopt $f_{\rm BLR}=1.3$ to calculate the BH mass for 3C~273,
but acknowledge that there is a large amount of uncertainty in the value
of this parameter (we do not include the uncertainty of $f_{\rm BLR}$ in our mass calculation).
Based on the measurements of $\tau_{\rm H\beta} = 146.8_{-12.1}^{+8.3}$
days (in the rest frame) and ${\rm FWHM_{H\beta}}=3314.1\pm59.3$ km/s,
the BH mass of 3C~273 is $\bhm = 4.1_{-0.4}^{+0.3} \times 10^8 M_{\odot}$. 

The relativistic jet is found to have an inclination of $10^{\circ}$ 
with a Lorentz factor of $\Gamma\approx 11$
\citep{Davis1991,Abraham1999}, which is nicely agreeable with the 
measurements ($i=12^{\circ}\pm 2^{\circ}$) of the GRAVITY (see Section \ref{sec:previous_GRAVITY}), showing that the BLR and the jet are perpendicular to
each other \citep{Sturm2018}. Since $\Theta_{\rm BLR}\approx 45^{\circ}$, the inclination ($12^{\circ}$) 
implies that the present estimate of SMBH mass is less affected by inclinations since the normalization
factor $f_{\rm BLR}\propto \left[(H_{\rm BLR}/R)^2+\sin^2i\right]^{-1}$ is dominated by the geometric factor
of $H_{\rm BLR}/R$ (i.e., $H_{\rm BLR}/R=\tan \Theta_{\rm BLR}$, where $H_{\rm BLR}$ is the half-thickness
of the BLR).

The GRAVITY observations provide an SMBH mass of $\bhm=(2.6\pm 1.1)\times 10^8\sunm$ \citep{Sturm2018},
which is consistent with the present results within the uncertainties.
We would like to point out that the most uncertainties of SMBH masses from Eq. (\ref{bhm}) are due 
to $f_{\rm BLR}$. 
Detailed modelling BLR through Markov Chain Monte Carlo (MCMC) simulations can be done for the
mass as well as the BLR geometry \citep{Pancoast2011,Pancoast2014a,Li2013,Li2018} independent of
this factor. Application of the MCMC technique
\footnote{The software developed by our SEAMBH team is publicly available at 
https://github.com/LiyrAstroph/BRAINS. It includes a designed model for BLRs with shadowed 
and unshadowed zones arising from the self-shadowing effects of slim accretion disks in 
super-Eddington accreting AGNs.}
%
%
%The software developed by our SEAMBH team is publicly available 
%at \url{https://github.com/LiyrAstroph/BRAINS}. It includes a designed model for  BLRs with shadowed and unshadowed %zones arising from the self-shadowing effects of slim accretion disks in super-Eddington accreting AGNs.}
% \footnote{The MCMC softeware independently developed by the SEAMBH
% team is publicly available at
% \url{https://github.com/LiyrAstroph/BRAINS}. It is designed for BLR with shadowed 
% and unshadowed zones determined by the self-shadowing effects of slim accretion disks 
% in super-Eddington accreting AGNs.} 
to 3C 273 will be carried 
out separately (Li et al. 2019 in preparation).

%\subsection{Accretion rates}
We use the equation for the dimensionless accretion rate of
\begin{equation}
\dot{\mathscr{M}}_{\bullet} = 20.1\,\left(\frac{\ell_{44}}{\cos i}\right)^{3/2}m_7^{-2},
\end{equation}
\cite[e.g.,][]{Du2014} to find $\dot{\mathscr{M}}_{\bullet}=\dot{M}_{\bullet}/L_{\rm Edd}c^{-2}\approx 9.6$
in 3C~273, where $L_{\rm Edd}$ is the Eddington luminosity,  $i (=12^{\circ})$ is the disk inclination angle,
$\ell_{44}=L_{5100}/10^{44}{\rm erg~s^{-1}}=84.3$, and
$m_7=\bhm/10^7\sunm=41.0$. Using the bolometric luminosity for 3C~273 of 
$L_{\rm bol}\approx 1.3\times 10^{47}{\rm erg~s^{-1}}$ determined by \citet{Shang2005} (The luminosity has 
been transformed to the value corresponding to the cosmological parameters used in this paper), we find that 
$L_{\rm bol}/L_{\rm Edd}\approx 2.4$, indicating that the BH is
accreting in this quasar at a mildly super-Eddington rate.

\begin{deluxetable}{lcccc}
	\tabletypesize{\footnotesize}
		\tablecaption{Comparison of the RM-campaigns\label{RM-campaigns}}
		\tablehead{
			\multirow{2}{*}{} &
			\multicolumn{2}{c}{Cadence} &
			\multirow{2}{*}{Length} &
			\multirow{2}{*}{$N_{\rm spec}$} \\ \cline{2-3}
			\colhead{} &
			\colhead{Obs. season} &
			\colhead{Entire} & (start$-$end yr)\\
			& (days) & (days) & (days) & 
		}
		\startdata
		\citet{Kaspi2000} & 40.3 & 66.8 & 2606 (1991-98)& 39 \\
		This campaign & 7.4 & 12.5 & 3400 (2008-18)& 296 \\
		\enddata
	\tablecomments{\footnotesize 
	``Obs. season" cadence means that it is averaged only during observable seasons and
	the entire one does that it is averaged for the entire campaign. Length refers to the total
	period from the start to end. $N_{\rm spec}$ is the number of spectra taken from the corresponding
	campaigns.}
	\end{deluxetable}

\section{discussion}
	\subsection{Comparing with the previous and the GRAVITY results}\label{sec:previous_GRAVITY}
	Before the present work, \citet{Kaspi2000} made a mapping measurement of 3C~273 through a
	7-year campaign of the Bok telescope and the Wise 1m telescope.
	They obtained the centroid lags of reverberation of the Balmer lines,
	$(R_{\rm H\alpha},R_{\rm H\beta},R_{\rm H\gamma})=(514_{-64}^{+65},382_{-96}^{+117},307_{-86}^{+57})$ltd, 
	respectively, that are twice more as long as the present results. But the error bars of the present 
	measurements are much smaller than those given by \cite{Kaspi2000}.
	\cite{Peterson2004} revisited the data of \cite{Kaspi2000} and found 
	$R_{\rm H\beta}=306.8_{-90.9}^{+68.5}$ltd, it is still double the present measured result in this paper.
	A brief comparison of the two campaigns is given by
	Table \ref{RM-campaigns}, showing that the present campaign has a much higher cadence than
	\cite{Kaspi2000}. Considering the BLR dynamical timescales of 
	$\tau_{\rm BLR}\approx R_{\rm BLR}/V_{\rm FWHM}=37.4\,R_{150}V_{3300}^{-1}$\,yrs,
	the BLR cannot change by a factor of 2 in the last 27 yrs (since 1991), where $R_{150}=R_{\rm BLR}/150$ltd 
	and $V_{3300}=V_{\rm FWHM}/3300\kms$. 
	Moreover, variations of the 5100\AA\, luminosity are
	about $\Delta L/L_{5100}\lesssim 1/4$ during the two campaigns and thus the difference of H$\beta$ lags 
	measured by the campaigns is not caused by luminosity changes. The present measurements are more robust 
	and reliable.
	
	Very recently\footnote{The 1st version of the paper appeared on November 9, 2018, and the GRAVITY 
	paper published on November 29, 2018.}, the GRAVITY instrument installed on the Very Large Telescope 
	Interferometer (VLTI) found a flatten disk 
	structure of the BLR in 3C 273 with a half-opening angle of $\Theta_{\rm BLR}={45^{\circ}}_{-6}^{+9}$ 
	and an emissivity-averaged radius of the Paschen-$\alpha$ emission region $R_{\rm Pa\alpha}=145\pm 35$ltd  
	\citep{Sturm2018}. It is exciting to find that $R_{\rm Pa\alpha}=R_{\rm H\beta}=R_{\rm H\gamma}$ holds 
	within their error bars as listed in Table \ref{table:timelag}. This lends a robust demonstration that 
	RM measurements are very reliable to spatially resolve compact regions in time domain through small 
	telescopes. 
	
	Moreover, the opening angle of the geometrically thick BLR in 3C 273 might agree with that
	of the dusty torus. This
	is helpful to understand the hypothesis that the BLR-clouds are supplied by 
	tidally captured clumps from the torus \citep{Wang2017}. In principle, the torus
	opening angles can be estimated by infrared emission as the reprocessed emission of optical and UV 
	luminosities \citep{Wang2005,Cao2005,Maiolino2007}, but the infrared emissions in 3C 273 could be seriously 
	contaminated by non-thermal emission from the jet.
	So it's hard to estimate the opening angle of the torus in this way. 
	Fortunately, the \oiii\, luminosity is
	$L_{\rm [O\,III]}\approx 1.5\times 10^{43}\ergs$ from this campaign, allowing an estimate of the torus 
	half-opening angle of $\Theta_{\rm torus}\approx 30^{\circ}\pm 10^{\circ}$ from the correlation 
	between type 2 quasar ratio and \oiii\, luminosity 
	\citep[i.e., the receding torus model, e.g. Fig. 12 in][]{Reyes2008}. This shows potential evidence for the hypothesis of the physical connection
	between the BLR and the torus. The GRAVITY observation 
	of the torus in 3C 273 in the future, as done in NGC 1068 \citep{Jaffe2004}, will directly measure the torus geometry
	to check if the geometrically-thick BLR matches that of the dusty torus. 
	This will
	observationally justify the hypothesis of the BLR origin from the failed outflows of the outer part accretion
	disk \citep{Czerny2011} or from torus \citep{Wang2017}. 

\subsection{Host galaxy}\label{sec:dis}
\begin{figure*}[htbp]
	\centering
	\includegraphics[trim=0 10 0 15,width=1.0\textwidth]{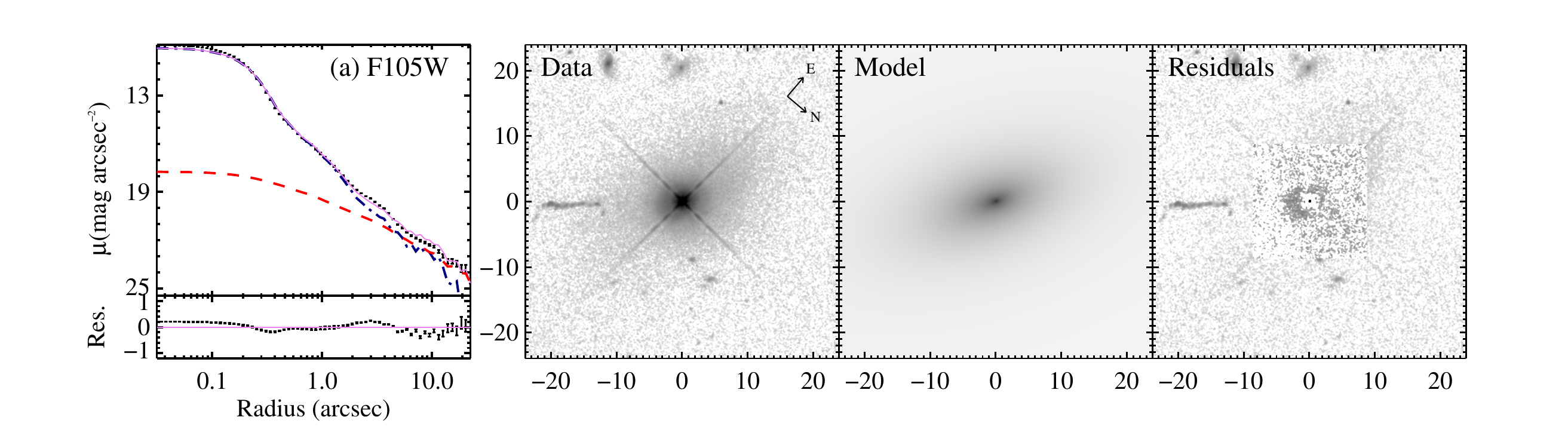}%{3c273_n4_F105W.pdf}
	
	\includegraphics[trim=0 10 0 15,width=1.0\textwidth]{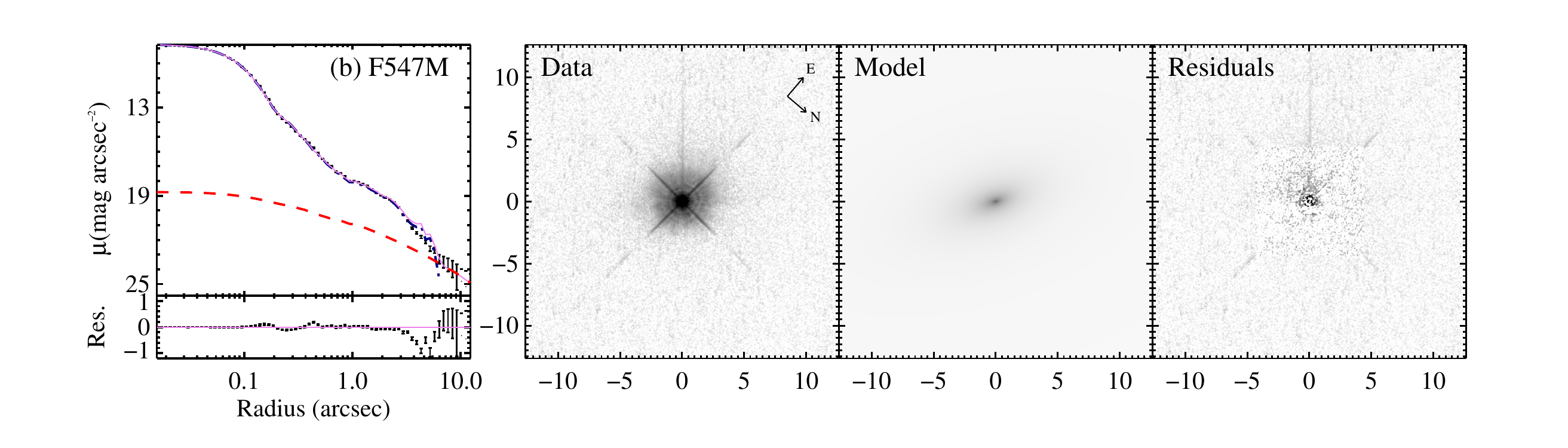}%{3c273_n4_F547M.pdf}
	\caption{\footnotesize GALFIT decomposition for 3C~273. The upper panels
	show the fitting result in F105W filter. The lower panels show the
	fitting result in F547M filter. In each row, 1D profile, 2D image
	of the original data, the best-fit model for the host (the AGN is excluded
	to better highlight the host) and the residuals images are displayed.
	The 1D azimuthally averaged profile shows the original data (black dots),
	the best fit (violet solid line), and the subcomponents (PSF in dark blue
	dot-dashed line and bulge in red dashed lines). The axes of the images
	are in units of arcseconds. All images are on an asinh stretch.
	}\label{Fig:galfit}
\end{figure*}
%
%\subsection{Stellar mass of the host galaxy of 3C~273}\label{sec:stellarmass}
\subsubsection{Data}
AGNs show strong correlations with their host galaxies, for example, 
the BH mass is strongly correlated with the mass of the galactic bulge of the host
\citep[e.g.,][]{Magorrian1998,Kormendy2013}. To derive the bulge mass in
3C~273, we analyze its \hst/WFC3 images\footnote{There is another archive data
of \hst\ observation led by \cite{Bentz2009}, but only provided one color
image. We revisited her data and obtained similar results with \cite{Bentz2009}
to remove host contaminations. However,
we need two-color images of \hst\ observations for stellar mass in this paper.}, 
which were observed on
March 17, 2013 (GO-12903, PI: Luis C. Ho) with the F547M filter in the UVIS
channel for 360 sec and with F105W filter in the IR channel for 147 sec.
Two additional short F547M exposures for 18 sec and 6 sec were taken to
warrant against saturation of the nucleus in the long exposure. These two
filters were selected to mimic the $B$ and $I$ bands in the rest
frame of 3C~273. To better sample the point-spread function (PSF), the long
UVIS observation was taken with a three-point linear dither pattern,
while the IR observation was taken with the four-point box dither pattern.
To avoid overheads due to buffer dump, we employed the UVIS2-M1K1C-SUB
1k$\times$1k subarray for the UVIS channel and the IRSUB512 subarray for the
IR channel, which results in a restricted field-of-view of 40$\times$40
arcsec and 67$\times$67 arcsec, respectively.

Because of the sparseness of field stars in the vicinity of 3C~273, the
observations were conducted in GYRO mode, leading to a
considerable degradation of the PSF of the dither-combined image generated
from the standard data reduction pipeline. Instead, we use the DrizzlePac
task AstroDrizzle (v1.1.16) to correct the geometric distortion,
align the sub-exposures, perform sky subtraction, remove cosmic rays, and,
finally combine the different exposures. The core of the AGN was severely
saturated in the long F547M exposure, and it was replaced with an
appropriately scaled version of the 6 sec short exposure. 
Because the PSF of \hst/WFC3 is under-sampled, we broaden the science 
image and the PSF by a Gaussian kernel such that the final images are 
Nyquist-sampled \citep{Kim2008a}. This largely removes the sub-pixel mismatch 
of the core of the PSF.

\begin{deluxetable*}{crcccccccccc}[ht]
	\tabletypesize{\footnotesize}
		\tablecaption{Fitted Model Parameters for \hst\ Images\label{tab:galfit}}
		\tablewidth{0pc}
		\tablehead{
		\colhead{Filter} &
		\colhead{$m_{\rm nuc}$} &
		\colhead{$m_{\rm host}$} &
		\colhead{$R_e$} &
		\colhead{$n$} &
		\colhead{$\mu_e$} &
		\colhead{RMS.} &
		\colhead{$\chi^2/\nu$} &
		\colhead{${a}_{1}$} \\
		\colhead{ } &
		\colhead{(mag) } &
		\colhead{(mag) } &
		\colhead{(\asec) } &
		\colhead{} &
		\colhead{(mag arcsec$^{-2}$)} &
		\colhead{} &
		\colhead{}   \\
		\colhead{ (1)} &
		\colhead{ (2)} &
		\colhead{ (3) } &
		\colhead{ (4) } &
		\colhead{ (5) } &
		\colhead{ (6)} &
		\colhead{ (7)} &
		\colhead{ (8)} &
		\colhead{ (9)} 
		}
		\startdata
		F105W  &\h $12.12 \pm 0.03$ &\h $14.87\pm 0.60$ &\h  $11.87\pm 5.76$ &\h $4 \pm 1$
		&\h $23.63 \pm 0.80$ 
		&\h 0.19 &\h 4.44   &\h  $-0.09 \pm 0.07$  \nl
		F547M  &\h $12.69\pm 0.01$ &\h $16.43 \pm 0.71$ &\h  $11.87\pm 5.76$ &\h $4 \pm 1$
		&\h $25.19 \pm 0.67$ 
		&\h 0.22 &\h 0.77   &\h  $-0.09 \pm 0.07$  	\\
		\enddata
		\tablecomments{\footnotesize
		Col. (1): \hst\ filter.
		Col. (2): Apparent nuclear magnitude in the observed filter.
		Col. (3): Apparent host magnitude fitted by a \sersic\ component in the observed filter.
		Col. (4): Effective radius of the host galaxy.
		Col. (5): \ser\ index for the host galaxy.
		Col. (6): Surface brightness of the host galaxy at the effective radius.
		Col. (7): RMS of the 1D profile residual in the source-dominated region.
		Col. (8): $\chi^2/\nu$ for the fit.
		Col. (9): Amplitude of the first Fourier mode, which is a measure of the lopsidedness of the galaxy.
		}
\end{deluxetable*}
\subsubsection{Image decomposition}
The F105W filter image for 3C~273 shows a dominant nucleus and a
host galaxy with no clear disk component. The F547M filter image is
considerably shallower. To extract quantitative measurements of the bulge,
we use the program GALFIT \citep{Peng2002,Peng2010} to fit two-dimensional surface
brightness distributions to the \hst\ images. A crucial ingredient is the
PSF, which will have a strong effect on the brightness of the active
nucleus. Unfortunately, no suitable bright star is available to be
used as the PSF within the limited field-of-view of the subarray WFC3
images. Instead, we generated a high-S/N empirical PSF by combining a large
number (24 for F105W and 12 for F547M) of bright, isolated, unsaturated
stars observed in other WFC3 programs. Extensive tests, consisting of 
fits to isolated bright stars, indicate that our stacked empirical PSF is far 
superior to synthetic  PSFs generated from the TinyTim program 
\citep{Krist2008}, and it has higher S/N than the PSFs of individual stars. The 
reduced chi-squared of the fits are $\sim$3 times larger for the TinyTim 
synthetic PSF.  Comparison of empirical PSFs observed from different programs 
indicate that the WFC3 PSF does not vary significantly with time ($<$ 10\%).

We first obtain the best global fit on the deeper F105W
image, whose redder wavelength coverage is more sensitive to the host. 
Models consisting of two components are adopted, with a point source 
(represented by the PSF) for the nucleus, and the bulge parameterized by 
a {S\'{e}rsic} function \citep{sersic1968} with index $n$. Models 
with $n \approx 3-5$ give the best fitting residuals.  We adopt 
$n=4$ as the best model and include the difference between the $n = 3$ and 
$n = 5$ model into the uncertainty. Figure \ref{Fig:galfit} summarizes the model 
fits in F105W and F547M filter for the $n=4$ model. We use the $m = 1$ Fourier 
mode, which is sensitive to lopsidedness, to gauge the degree of global asymmetry of
the galaxy \citep[see, e.g.,][]{Kim2008b,Kim2017}. Since the host is
considerably weaker in the F547M image, it is fit by keeping the structural
parameters fixed to the values obtained from the F105W model, solving only
for the magnitudes of the nucleus and host. 

The best-fit parameters are given in Table \ref{tab:galfit}. For bright
type 1 AGNs, such as 3C~273, the uncertainties are dominated by 
errors in the PSF adopted for the nucleus. We estimate the 
uncertainties in the PSF
by generating variants of the empirical PSF by combining different subsets
of stars, and then repeating the fit. Subtraction of the background can 
also affect the fit. We subtract $\pm1 \sigma$ of the 
background level from the image, and then take the mean of the 
differences relative to the fiducial value of the sky as the 
uncertainty.  The uncertainties caused by the PSF, background subtraction, and 
the {S\'{e}rsic} index $n$ are all taken into account in the final error estimation.

\subsubsection{Stellar mass}\label{sec:stellarmass}
The GALFIT decomposition of the \hst\ images yields F547M and F105W magnitudes
for the bulge of the host galaxy of 3C~273. We convert these
\hst-based magnitudes into intrinsic, rest frame $I$-band magnitude and
$B-I$ color, from which we can estimate the stellar mass following the
prescriptions of \cite{Bell2001}. To estimate the $K$-correction, we use
\citet{Bruzual2003} models with solar metallicity, a \cite{Chabrier2003}
stellar initial mass function (IMF), and an exponentially decreasing star
formation history with a star formation timescale of 0.6 Gyr to generate a
series of template spectra with ages spanning 1 to 12 Gyr. After
accounting for Galactic extinction and redshift, we convolve the spectra
with the response functions of the \hst\ filters to generate synthetic
F547M and F105W magnitudes.

The observed host color is $\mathrm{F547M - F105W} = 1.6\pm 0.9$.  This color
is best matched with a stellar population template  of ${2.5}$ Gyr (the
templates of 1 Gyr and 12 Gry for the lower and upper limits),  resulting in
$M_I = -24.3 \pm 0.6$ and  $B-I$ = $1.5 \pm 1.0$. Following \citet{Bell2001}
and the correction of \citet{Longhetti2009}, we derive a stellar mass of $M_*
= 10^{11.3 \pm 0.7} M_\odot$, assuming solar metallicity and a Chabrier IMF.
The ratio of BH mass and spheroid is about $2.0\times 10^{-3}$,
which agrees with the Magorrian relation \citep[e.g.,][]{Kormendy2013}.

\section{summary}\label{sec:sum}

We present the results from a reverberation mapping campaign of
3C~273
from Nov 2008 to Mar 2018. 
Time lags relative to the observed continuum variations of the
	active nucleus of several emission lines were successfully detected, and we
	measure the velocity-resolved lags for the \hbeta\ line.
				
\begin{itemize}
\item The time lags of the \hbeta\ and H$\gamma$ emission lines are
			 $(\tau_{\rm H\beta},\tau_{\rm H\gamma})=(146.8_{-12.1}^{+8.3},146.5_{-9.7}^{+8.8})$ 
			 days in the rest frame, which are very similar to each other.

\item   The \feii\ lines have a lag of $\tau_{\rm Fe}=322.0_{-57.9}^{+55.5}$
			days in the rest frame, which follows the
			$\tau_{\rm Fe}/\tau_{\rm H\beta}-F_{\rm Fe}/F_{\rm H\beta}$
			correlation found in \citet{Hu2015}. The \feii\ and the Balmer line regions follow
			the virialized relation of $\tau\propto V_{\rm FWHM}^{-2}$, showing the
			stratified structures in space.

\item   The velocity-resolved lag measurements of the \hbeta\
			line show a complex structure. The lags are longer at small velocities 
			and shorter at high velocities, which is the signature of a rotation-dominated 
			disk. In addition, the blue wing shows longer lags than 
			those at red velocities. This may be explained by inflowing radial motions in 
			the \hbeta-emitting gas or some other special BLR kinematics.

\item   Along with the UV lines, we find that 3C~273 has a
			stratified structure in its BLR, with higher-ionization lines arising from inner regions but
            lower-ionization lines from the outer part. The UV lines 
			deviate significantly from the ${\rm FWHM}\propto \tau^{-0.5}$ relation.

\item   Adopting a virial factor of $f_{\rm BLR} = 1.3$, 3C~273 has a BH mass of 
	        $\bhm = 4.1_{-0.4}^{+0.3} \times 10^8 M_{\odot}$
			and is accreting with a rate of $9.3\,L_{\rm Edd}\,c^{-2}$,  indicating that
			the black hole is undergoing intermediate super-Eddington accretion. 
			
\item   From decomposition of its \hst\ images, we obtain a host
			stellar mass of $M_* = 10^{11.3\pm0.7} M_\odot$.  We find that a ratio of BH mass
			and host spheroid ($2\times 10^{-3}$) follows the Magorrian relation. 
\end{itemize}
				
 Considering that the geometrically thick BLR of 3C 273 detected by the GRAVITY is roughly 
consistent with the torus obtained by the receding statistic, we expect to compare with torus 
structure spatially resolved by the VLTI in order to reveal the origin of BLR clouds, which is 
suggested to be supplied by the central black hole tidal capture of clumps from the 
torus \citep{Wang2017}.

We thank an anonymous referee for a helpful report. We acknowledge the support by
National Key R\&D Program of China (grants 2016YFA0400701 and
2016YFA0400702), by NSFC through grants {NSFC-11873048, -11833008, 
-11573026, -11473002, -11721303, -11773029, -11833008, -11690024}, and by Grant No. QYZDJ-SSW-SLH007 from the Key Research Program
of Frontier Sciences, CAS, by the Strategic Priority Research Program of the Chinese Academy of Sciences grant No.XDB23010400. Data from the Steward Observatory
spectropolarimetric monitoring project were used. This program is supported
by Fermi Guest Investigator grants NNX08AW56G, NNX09AU10G, NNX12AO93G,
and NNX15AU81G. We also acknowledge the support of the staff of the Lijiang 2.4m telescope.
Funding for the telescope has been provided by CAS and the People's
Government of Yunnan Province.

\appendix

\begin{figure*}[ht]
	\begin{center}
		\includegraphics[width=0.8\textwidth]{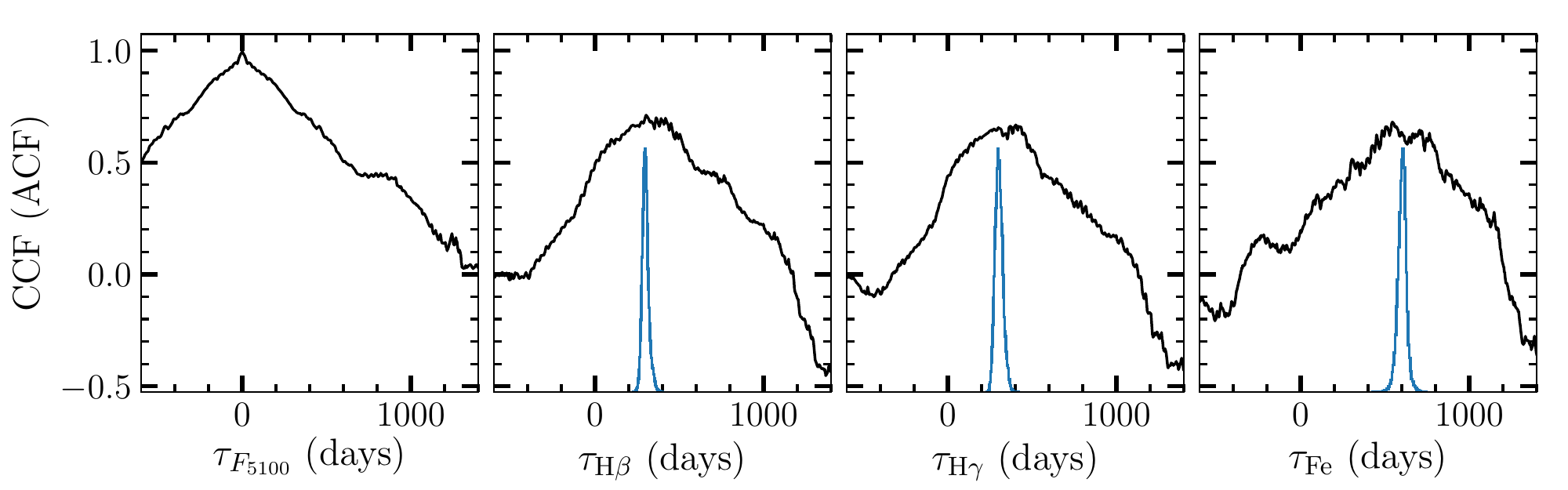}%{fig_lightcurve.pdf}
	\end{center}
	\caption{\footnotesize
	Light curves and CCFs without detrending. The meanings of the panels, symbols, lines, 
	and units are the same as in Figure \ref{fig:lightcurve}.
		\label{fig:lightcurve_nodetrend}}
\end{figure*}

\begin{figure*}[ht]
	\centering
	\includegraphics[width=0.5\textwidth]{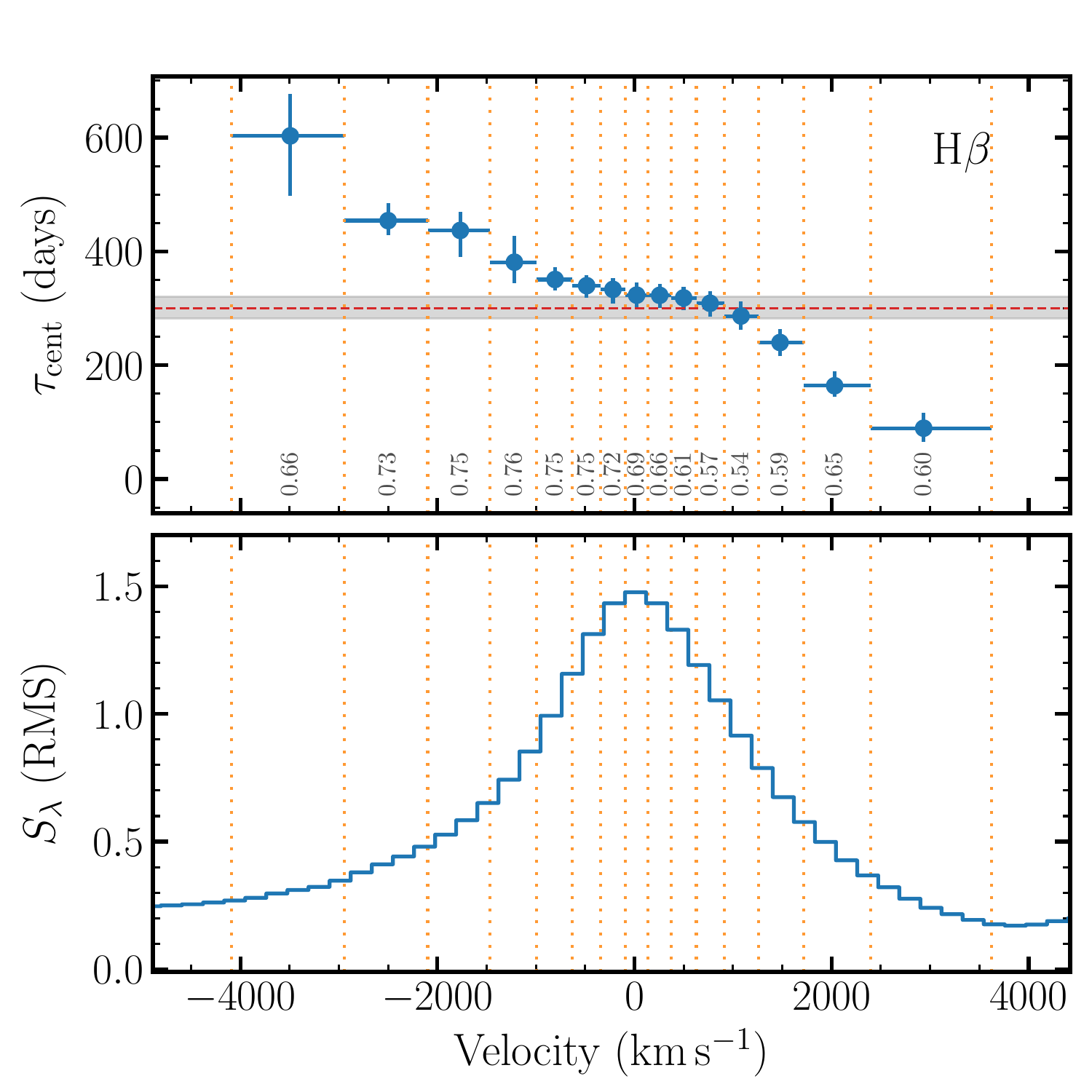}%{fig_velo_beta.pdf}
	\caption{\footnotesize
		Velocity-resolved time lags of the \hbeta\ line without detrending
		in the observed frame. The meanings of the panels, symbols, lines, 
	and units are the same as in Figure  \ref{fig:velo}.
	}\label{fig:velo_nodetrend}
\end{figure*}

For completeness, we also show the results of the time-series analysis and the
velocity-resolved lag measurement without detrending in Figures
\ref{fig:lightcurve_nodetrend} and \ref{fig:velo_nodetrend}. The
time lags of the \hbeta, \hgamma, and \feii\  lines are listed in Table
\ref{table:timelag_nodetrend}. The long-term variations in 5100\AA\ light
curve (i.e., the long-term trending) bias the lag measurements and  the velocity-resolved results
significantly. Comparing with Table \ref{table:timelag}, we find that CCF correlations are also 
stronger than the results without detrending (see Table \ref{table:timelag_nodetrend}). We adopt 
the detrended results given in Table \ref{table:timelag} of the main text.

\begin{deluxetable*}{lccc}%[!h]
\tabletypesize{\footnotesize}
	\tablecaption{Lags without detrending\label{table:timelag_nodetrend}}
	\tablehead{
		\multirow{2}{*}{Lines} &
		\multirow{2}{*}{$r_{\rm max}$} &
		\multicolumn{2}{c}{lags (days)}\\ \cline{3-4}
		\colhead{} &
		\colhead{} &
		\colhead{observed frame} &
		\colhead{rest frame}
	}
	\startdata
	\hbeta  & 0.71 & $298.1^{+19.3}_{-16.8}$ & $257.4^{+16.6}_{-14.5}$ \\
	\hgamma & 0.67 & $302.5^{+23.5}_{-21.8}$ & $261.2^{+20.3}_{-18.9}$ \\
	\feii   & 0.68 & $603.0^{+21.8}_{-25.6}$ & $520.6^{+18.8}_{-22.1}$ \\
	\enddata
\end{deluxetable*}

\end{document}